\documentclass[12pt]{article}
\usepackage{latexsym}
\usepackage{amssymb}
\usepackage{graphicx}
\usepackage{multibox}
\usepackage{amsmath,amsfonts}

\topskip \textwidth 480pt

\def\theequation{\thesection.\arabic{equation}}

\newcommand{\be}{\begin{equation}}
\newcommand{\ee}{\end{equation}}

\def\bea{\begin{eqnarray}}
\def\eea{\end{eqnarray}}
\def\nn{\nonumber}
\def\um{{\underline{m}}}
\def\un{{\underline{n}}}

\begin{document}
\begin{titlepage}
\title{
{\bf BLG--motivated Lagrangian formulation for the chiral two-form
gauge field in $D=6$ and M5--branes}
\medskip
\medskip
\author{Paolo Pasti\,$^{*\dagger}$, Igor Samsonov\,$^{\dagger\ddag}$,
Dmitri~Sorokin\,$^{\dagger}$ and Mario Tonin\,$^{*\dagger}$
~\\
~\\
{\it $^*$  Dipartimento di Fisica ``Galileo Galilei",} ~\\
 {\it  Universit\'a degli Studi di Padova} ~\\
~\\
{\it $^\dagger$ Istituto Nazionale di Fisica Nucleare, Sezione di
Padova,}
~\\
{\it via F. Marzolo 8, 35131 Padova, Italia}
~\\
~\\
{\it $^\ddag$ Laboratory of Mathematical Physics,}
~\\
{\it Tomsk Polytechnic University, 634050 Tomsk, Russia}} }

\date{}
\maketitle

\begin{abstract}
\noindent
We reveal non--manifest gauge and $SO(1,5)$ Lorentz symmetries in
the Lagrangian description of a  six--dimensional free chiral field
derived from the Bagger--Lambert--Gustavsson model in
arXiv:0804.3629 and make this formulation covariant with the use of
a triplet of auxiliary scalar fields. We consider the coupling of
this self--dual construction to gravity and its supersymmetrization.
In the case of the non--linear model of arXiv:0805.2898 we solve the
equations of motion of the gauge field, prove that its non--linear
field strength is self--dual and find a gauge--covariant form of the
non--linear action. Issues of the relation of this model to the
known formulations of the M5--brane worldvolume theory are
discussed.

\end{abstract}
~
\thispagestyle{empty}
\end{titlepage}
\tableofcontents

\section{Introduction}
The problem of the Lagrangian formulation of the theory of
self--dual or in general duality--symmetric fields,
\emph{i.e.} fields whose strengths are subject to a duality
condition, has attracted a great deal of attention for decades. A
classical physical example, the duality symmetry between electric
and magnetic fields of free Maxwell equations, inspired Dirac to
promote it to the gauge theory of electrically and magnetically
charged particles by introducing the magnetic monopoles
\cite{Dirac}. Since then duality--symmetric fields appeared and have played an
important role in many field theories, in particular, in String
Theory and M--theory. The gauge fields whose field strength is
self--dual are often called chiral (p--form) fields. In space--times
of Lorentz signature such fields exist if $p=2k$ $(k=0,1,\ldots)$
and the space--time dimension is $D=2(p+1)$.

Main problems of the Lagrangian formulation of the
duality--symmetric and, in particular, the chiral fields are i) to
construct an action whose variation would produce the
\emph{first--order} duality condition on the field strengths as a
consequence of dynamical equations of motion; ii) to find a
manifestly Lorentz--covariant form of such an action, which is of a
great help for studying a (non--linear) coupling of
duality--symmetric fields to gravity and other fields in the theory;
iii) to quantize such a theory.

The first two (classical) problems have been solved in a number of
papers using different (classically equivalent) approaches. It has
been realized that it is not possible to construct manifestly
duality--symmetric and Lorentz--covariant actions without using
auxiliary fields. In various space--time dimensions non--manifestly
Lorentz covariant duality--symmetric actions were constructed and
studied in \cite{Zwan}--\cite{SS}, see also \cite{Belov:2006jd} for
more recent developments based on a holographic formulation of
self-dual theory. It is known that in these models the
Lorentz--invariance gets restored at the level of the equations of
motion (\emph{i.e.} when the duality relation holds) and is actually
a somewhat modified non--manifest symmetry of the action (see  e.g.
\cite{SS}).

To make the Lorentz invariance of the duality--symmetric action
manifest, in particular that of the chiral field action, one should
introduce auxiliary fields. In different formulations their amount
vary from infinity
\cite{mac,wot,ben,berkovits} to a few
\cite{Siegel,KM} or even one \cite{pst,PST1}. The relation between
different non--covariant and covariant formulations was studied
\emph{e.g.} in \cite{PST1,Maznytsia:1998xw,Miao:2000fn}. The
quantization of duality--symmetric and chiral gauge fields (which is
a subtle and highly non--trivial problem, especially in
topologically non--trivial backgrounds) has also been intensively
studied, see
\emph{e.g.} \cite{Zwan,Belov:2006jd,mac,wot,ben,berkovits,anomaly}--\cite{Berman:2007bv}
and references therein.

One more, non--covariant, Lagrangian formulation of a chiral 2--form
gauge field in six space--time dimensions was derived in
\cite{Ho1,Ho2} from a Bagger--Lambert--Gustavsson (BLG) model of
interacting Chern--Simons and matter fields in $D=3$
\cite{Bagger:2006sk,Gustavsson:2007vu}. This has been achieved
by promoting the non--Abelian gauge symmetry of the BLG model to the
infinite--dimensional local symmetry of volume preserving
diffeomorphisms in an internal 3--dimensional space, see also
\cite{Bandos:2008jv,Bandos:2008fr}. It was argued in
\cite{Ho1,Ho2} that when the initial $D=3$ space--time and the
3--dimensional internal space are treated as six--dimensional
space--time, such a model describes a non--linear effective field
theory on the worldvolume of a 5--brane of M--theory in a strong
$C_3$ gauge field background. Other aspects of the relation of the
M5--brane to the BLG model based on the 3--algebra associated with
volume preserving diffeomorphisms were considered \emph{e.g.} in
\cite{Bandos:2008fr,Bonelli:2008kh,Chu:2009iv}. In particular, the authors
of \cite{Bandos:2008fr} found a relation of the M5--brane action
\cite{PST2,M5S,Bergshoeff:1998vx}, in the limit of infinite M5--brane tension, to a
Carrollian limit of the BLG model in which the speed of light is
zero (which amounts to suppressing all spacial derivatives along the
M2--brane).

The aim of this paper is to discuss and clarify some issues of the
$D=6$ chiral field model of \cite{Ho1,Ho2} regarding its space--time
and gauge symmetries, and self--duality properties. We shall first
consider the free chiral field formulation of \cite{Ho1} and then
its non--linear generalization constructed in \cite{Ho2}. We shall
also compare this model with the original actions for the $D=6$
chiral 2--form gauge field \cite{HT,SS,PST1}, as well as with the
M5--brane action \cite{PST2,M5S} and equations of motion
\cite{Howe:1996yn,Howe:1997fb,Howe:1997vn,Bandos:1997gm,Cederwall:1997gg}.

In the free field case, we show that, like the actions of
\cite{HT,SS}, the quadratic chiral field action of \cite{Ho1}
possesses a non-manifest six--dimensional (modified) Lorentz
symmetry and can be covariantized, coupled to gravity and
supersymmetrized in a way similar to the approach of
\cite{pst,PST1}. However it differs from the original PST
formulation in the number of auxiliary fields required for making
the $D=6$ chiral field action of
\cite{Ho1} manifestly covariant. We show that the latter
requires three scalar fields, taking values in the 3--dimensional
representation of a $GL(3)$ group, while the formulation of
\cite{pst,PST1} makes use of a single auxiliary scalar field.
This is expected, since in the model of \cite{Ho1,Ho2} the six
space--time directions are subject to 3+3 splitting, instead of the
1+5 splitting of \cite{HT,SS,pst,PST1} and
\cite{PST2,M5S}.

We then consider the non--linear chiral field model of
\cite{Ho1,Ho2} neglecting its couplings to scalar and spinor matter fields.
By solving the non--linear field equations derived in
\cite{Ho2} we find an explicit form of gauge field strength
components that were missing in the formulation of
\cite{Ho2} and show that the complete $D=6$ field strength transforms
as a scalar field under volume preserving diffeomorphisms and
satisfies the complete set of Bianchi relations. We prove that the
general solution of the non-linear field equations results in the
Hodge self--duality of the $D=6$ non-linear gauge field strength,
thus confirming the assumption of
\cite{Ho2}. We also find that the action of the non--linear model can be
rewritten in a form that involves solely the components of the
chiral field strength and hence is covariant under the volume
preserving diffeomorphisms.

The paper is organized as follows. In Section 2 we recall the basic
properties of a free 2--form chiral field in six dimensional
space--time (Section \ref{free}), consider the structure of a
non--covariant action for the $D=6$ chiral gauge field a la
\cite{HT,SS} (Section \ref{nonc}) and  overview the covariant Lagrangian description of the chiral fields  proposed and
developed in
\cite{pst,PST1} (Section \ref{pst}). In Sections \ref{hohoho} and \ref{hohohoho} we consider the alternative
non--covariant formulation of \cite{Ho1,Ho2} at the free--field
level and reveal its hidden gauge and Lorentz symmetries. In
Sections
\ref{covho}--\ref{coupling} we propose its covariantization, coupling to gravity and
supersymmetrization along the lines of the approach of
\cite{pst,PST1}.
In Section \ref{nonl} we consider the non--linear generalization of
the alternative chiral field formulation and study its symmetry and
self--duality properties. In Section
\ref{M5} we briefly discuss issues of the relation of the model of
\cite{Ho1,Ho2} to the worldvolume theory of the M5--brane.

\section{Actions for the $D=6$ chiral field}\label{chiral}
\subsection{The antisymmetric 2--rank gauge field in
$D=6$}\label{free}
Let $R^{1,5}$ be a six-dimensional Minkowski space having the metric
$\eta_{\mu\nu}={\rm diag}(-1,1,1,1,1,1)$ and parametrized by
coordinates $x^\mu$ $(\mu=0,1,\ldots,5)$. Let $A_{\mu\nu}$ be a
two--rank antisymmetric tensor field with the field strength
\be
F_{\mu\nu\rho} =\partial_\mu A_{\nu\rho}+\partial_\nu A_{\rho\mu}
+\partial_\rho A_{\mu\nu}\,.
\label{F}
\ee
The field strength (\ref{F}) is invariant under the gauge
transformations
\footnote{We use the symmetrization and antisymmetrization of
indices with `strength one', i.e. with the normalization factor
$\frac{1}{n!}$.}
\be\label{gs}
\delta A_{\mu\nu}=2\partial_{[\mu}\lambda_{\nu]}(x)
=\partial_\mu\lambda_\nu -\partial_\nu\lambda_\mu\,.
\ee
These gauge transformations are reducible because of the residual
gauge invariance of the gauge parameter,
\be\label{gauges}
\delta\lambda_\mu=\partial_\mu\lambda(x)\,.
\ee

The classical action for this field is
\be\label{a2action}
S=-\frac1{4!} g^2\int d^6x\, F_{\mu\nu\rho}F^{\mu\nu\rho}\,,
\ee
where $g$ is a coupling constant of mass dimensionality, which we
shall put equal to one in what follows. The corresponding equation
of motion is
\be
\frac{\delta S}{\delta A_{\mu\nu}}
=\partial_\rho F^{\mu\nu\rho}=0\,.
\ee
By definition, the field (\ref{F}) satisfies the Bianchi identity
\be
\varepsilon^{\alpha\beta\gamma\delta\rho\sigma}
\partial_{\gamma}F_{\delta\rho\sigma}=0\,.
\ee
On the mass shell, such an antisymmetric tensor field $A_{\mu\nu}$
describes six degrees of freedom. This number can be reduced to
three if one imposes an additional, self--duality, condition
\be
F_{\mu\nu\rho} =\tilde F_{\mu\nu\rho}\,,
\label{selfd}
\ee
where
\be\label{dual}
\tilde F_{\mu\nu\rho}:=\frac16
\varepsilon_{\mu\nu\rho\alpha\beta\gamma}
F^{\alpha\beta\gamma}\,.
\ee

 The field $A_{\mu\nu}$ satisfying eq.\  (\ref{selfd}) is called
the chiral field.

A natural question is whether one can derive the first--order
self--duality condition (\ref{selfd}) from an action principle as an
equation of motion of $A_{\mu\nu}$.  The answer is positive, though
the construction is non--trivial, and the resulting action possess
peculiar properties to be reviewed in the next Section.

\subsection{Non--covariant action}\label{nonc}
Usually the actions for free bosonic fields are of a quadratic order
in their field strengths, like eq.\  (\ref{a2action}). So if, in
order to get a chiral field action, one tries to modify the action
(\ref{a2action}) with some other terms depending solely on
components of $F_{\mu\nu\rho}$, one gets the equations of motion
that are of the second order in derivatives. Thus, the chiral field
action should have a structure and symmetries which would allow one
to reduce the second order differential equations to the
first--order self--duality condition. Such actions have been found
for various types of chiral fields
\cite{Zwan}--\cite{SS} but they turn out to be non--manifestly
space--time invariant. In the $D=6$ case the self--dual action can
be written in the following form
\be\label{21}
S=-\frac1{4!}\int d^6x [F_{\lambda\mu\nu}F^{\lambda\mu\nu}
+3(F-\tilde F)_{0ij} (F-\tilde F)^{~ij}_0]\,, \qquad (i,j =
1,\ldots,5)\,.
\ee
It contains the ordinary kinetic term for $A_{\mu\nu}$, and the
second term which breaks manifest Lorentz invariance down to its
spatial subgroup $SO(5)$, since only the time components of
$(F-\tilde F)$ enter the action \footnote{Alternatively, but
equivalently, one might separate one spacial component from other
five and construct an $SO(1,4)$ invariant action similar to
(\ref{21}) but in which the sign of the second term is changed and
the time index 0 is replaced with a space index,
\emph{e.g.} 5. This choice is convenient
when performing the dimensional reduction of the $D=6$ theory to
$D=5$.}. However, it turns out that eq.\
\eqref{21} is (non--manifestly) invariant under modified space--time
transformations
\cite{FJ,SS} which (in the gauge $A_{0i}=0$ for the local symmetry (\ref{gs})) look as follows
\be\label{31}
\delta A_{ij}=x^0v^k\partial_kA_{ij}+x^kv^k\partial_0A_{ij}-
x^kv^k(F-\tilde F)_{0ij}\,.
\ee
The first two terms in \eqref{31} are standard Lorentz boosts with a
velocity $v_i$ which extend $SO(5)$ to $SO(1,5)$. The last term is a
non--conventional one, it vanishes when
\eqref{selfd} is satisfied, so that the transformations \eqref{31} reduce to the
conventional Lorentz boosts on the mass shell.

  From \eqref{21} one gets the $A_{\mu\nu}$ field equations, which
have the form of Bianchi identities
\be\label{em}
\varepsilon^{ijklm}\partial_{k}(F-\tilde F)_{lm0}=0\,.
\ee
Their general (topologically trivial) solution is
\be\label{5}
(F-\tilde F)_{ij0}=2\,\partial_{[i}\phi_{j]}(x)\,.
\ee
If the right hand side of \eqref{5} were zero, then
\be\label{sd0}
F_{ij0}=\tilde F_{ij0}=\frac{1}{6}\,\varepsilon_{ijklm}\,F^{klm}
\ee
and, hence, as one can easily check, the full covariant
self--duality condition is satisfied. And this is what we would like
to get. One could put the r.h.s. of \eqref{5} to zero if there is an
additional local symmetry of \eqref{21} for which
$\partial_{[i}\phi_{j]}=0$ is a gauge fixing condition. And there is
indeed such a symmetry \cite{SS} which acts on the components of
$A_{\mu\nu}$ as follows
\be\label{6}
\delta A_{0i}=\Phi_i(x),\qquad \delta A_{ij}=0, \qquad
\delta(F-\tilde F)_{ij0}=2\,\partial_{[i}\Phi_{j]}.
\ee
The existence of this symmetry is the reason why the quadratic
action describes the dynamics of the self--dual field $A_{\mu\nu}$
with twice less physical degrees of freedom than that of a
non--self--dual one. It also implies that the components $A_{0i}$
are pure gauge and enter the action only under a total derivative.
$A_{0i}$ can be thus put to zero directly in the action, which fixes
the gauge symmetry \eqref{6}. The action \eqref{21} then reduces to
\be\label{22}
S=-\frac1{4!}\int d^6x [2F_{ijk}F^{ijk}
+\varepsilon^{ijklm}\,F_{klm}\,
\partial_0A_{ij}]\,,
\qquad (i,j = 1,\ldots,5)\,.
\ee
Eq.\ \eqref{22} does not contain the $A_{0j}$ component of the
six--dimensional chiral field. Thus, on the mass shell, the role of
this component is taken
 by the ``integration" function
$\phi_j(x)$ of \eqref{5}, which appears upon solving the second
order field equation
\eqref{em}. We shall encounter the same feature in the alternative
formulation of \cite{Ho1}, but before describing the construction of
\cite{Ho1} let us first review a covariant Lagrangian description of
the chiral field proposed in \cite{pst,PST1}.

\subsection{Lorentz--covariant formulation}\label{pst}
The covariant formulation of \cite{pst,PST1} is constructed with the
use of a single auxiliary scalar field $a(x)$. The covariant
generalization of the action \eqref{21} for the $D=6$ self--dual
field looks as follows
\begin{equation}\label{fina}
S=-\frac1{4!}\int d^6x [F_{\lambda\mu\nu}F^{\lambda\mu\nu}
-{3\over{(\partial_\rho a\partial^\rho a)}}\partial^\mu
a(x)(F-\tilde F)_{\mu\lambda\sigma}
\,(F-\tilde F)^{\lambda\sigma\nu}\partial_\nu a(x)]\,.
\end{equation}
 In addition to standard gauge symmetry \eqref{gauges} of $A_{\mu\nu}(x)$
the covariant action (\ref{fina}) is invariant under two different
local transformations:
\begin{equation}\label{varm}
\delta A_{\mu\nu} =2\,\partial_{[\mu}a\,\Phi_{\nu]}(x)\,, \qquad
\delta a=0\,;
\end{equation}
\begin{equation}\label{a}
\delta a=\varphi(x)\,, \qquad
\delta A_{\mu\nu}=
{{\varphi(x)}\over{(\partial a)^2}}(F-\tilde F)_{\mu\nu\rho}
\partial^\rho a\,.
\end{equation}
The transformations \eqref{varm} are a covariant counterpart of
\eqref{6} and play the same role as the latter in deriving the
self--duality condition \eqref{selfd}.

Local symmetry \eqref{a} ensures the auxiliary nature of the field
$a(x)$ required for keeping the space--time covariance of the action
manifest \cite{pst}. An admissible gauge fixing condition for this
symmetry is
\be\label{gfa}
\frac{\partial_\mu a(x)}{\sqrt{-\partial_\nu\,a\,\partial^\nu\,a}}=\delta^0_\mu\,.
\ee
In this gauge the action \eqref{fina} reduces to \eqref{21}. The
modified space--time transformations \eqref{31}, which preserve the
gauge
\eqref{gfa} arise as a combination of the Lorentz boost and the
transformation \eqref{a} with $\varphi=-v^ix^i$, $(i=1,2,3,4,5)$.

One may wonder whether by using the gauge transformation
\eqref{a} one can put the field $a(x)$ to zero.
This is indeed possible if one takes into account the subtlety that
by imposing such a gauge fixing one should handle a singularity in
the action
\eqref{fina} in such a way that the ratio
${\partial_\mu a\,\partial^\nu a}/{\partial_\rho a\partial^\rho a}$
remains finite. This can be achieved by first imposing the gauge
fixing condition $a(x)=\epsilon\,x^\mu\,n_\mu$, where $n_\mu$ is a
constant time--like vector $n^2=-1$ and then sending the constant
parameter $\epsilon$ to zero. As one can see, such a limit is
compatible with the gauge choice \eqref{gfa} with
$n_\mu=\delta^0_\mu$.

For further analysis it is useful to note that the auxiliary field
$a(x)$ enters the action \eqref{fina} only through the combination
which forms a projector matrix of rank one
\be\label{pa}
P_{\mu}{}^{\nu}=\frac{1}{\partial_\rho a\partial^\rho
a}\,\partial_\mu a\,\partial^\nu a\,,\qquad
P_{\mu}{}^{\rho}\,P_{\rho}{}^{\nu}=P_{\mu}{}^{\nu}\,.
\ee
Then the action \eqref{fina} takes the form
\begin{equation}\label{finap}
S=-\frac1{4!}\int d^6x [F_{\lambda\mu\nu}F^{\lambda\mu\nu}
-3(F-\tilde F)^{\lambda\sigma\nu}
\,P_{\nu}{}^{\mu}\,(F-\tilde F)_{\mu\lambda\sigma}
]\,.
\end{equation}
It produces the following Lorentz--covariant counterpart of the
self--duality condition
\eqref{sd0}
\be\label{sd0l}
{{1}\over{\sqrt{-(\partial a)^2}}}\,F_{\mu\nu\rho}\partial^\rho a =
{{1}\over{6\sqrt{-(\partial a)^2}}}\,
\partial^\rho
a\,\varepsilon_{\rho\mu\nu\lambda\sigma\tau}
\,F^{\lambda\sigma\tau}\equiv\tilde
F_{\mu\nu}.
\ee
As one can easily see, eq. \eqref{sd0l} is equivalent to the
self--duality condition \eqref{selfd}.

\setcounter{equation}0
\section{Free $D=6$ chiral gauge field from the BLG model}\label{hohoh}
\subsection{Non--covariant formulation}\label{hohoho}
A different non--covariant Lagrangian description of the $D=6$
chiral field was obtained in \cite{Ho1,Ho2} from a
Bagger--Lambert--Gustavsson (BLG) model
\cite{Bagger:2006sk,Gustavsson:2007vu} of interacting Chern--Simons
and matter fields in $D=3$ by promoting the gauge symmetry of the
BLG model to the infinite--dimensional local symmetry of volume
preserving diffeomorphisms of an internal 3--dimensional space. The
original 3--dimensional space--time (supposed to be a worldvolume of
coincident M2--branes) was assumed in \cite{Ho1,Ho2} to combine with
the 3--dimensional internal space and to form the 6--dimensional
worldvolume of a 5--brane carrying a 2--form chiral field. So in the
formulation of \cite{Ho1,Ho2} the $D=6$ Lorentz symmetry $SO(1,5)$
is (naturally) broken by the presence of membranes to $SO(1,2)\times
SO(3)$. In particular, the action for the free chiral field is
constructed with the use of components of $A_{\mu\nu}$ which are
split into $SO(1,2)\times SO(3)$ tensors and is thus an
$SO(1,2)\times SO(3)$ invariant counterpart of the $SO(5)$ (or
$SO(1,4)$) covariant chiral field Lagrangian of Subsection
\ref{nonc}.

We shall now briefly review this formulation for the case of the
free gauge field. The non--linear chiral field model of
\cite{Ho1,Ho2} will be discussed in Section \ref{nonl}.

With respect to the subgroup $SO(1,2)\times SO(3)$, the $SO(1,5)$
components of $A_{\mu\nu}$ split as follows
\be\label{sosplit}
A_{\mu\nu}=(A_{ab},\, A_{a\dot b},\, A_{\dot a\dot b})\,,
\ee
where the indices $a=(0,1,2)$ and $\dot a=(1,2,3)$, correspond,
respectively, to the $SO(1,2)$ and  $SO(3)$  subgroup of the full
$D=6$ Lorentz group. Each of the antisymmetric fields $A_{ab}$ and $A_{\dot
a\dot b}$ has three components, while $A_{a\dot b}$ has nine
components. The $D=6$ coordinates $x^\mu$ split into $x^a$ and
$x^{\dot a}$.

Only the components $A_{a\dot b}$ and $A_{\dot a\dot b}$ were used
in the construction of the chiral field Lagrangian of \cite{Ho1},
which has the form
\be\label{hoho}
L=-\frac{1}{4}\,F_{a\dot b \dot c}(F-\tilde f)^{a\dot b \dot
c}-\frac{1}{12}\,F_{\dot a\dot b\dot c}\,F^{\dot a\dot b\dot c}\,,
\ee
where
\be\label{fadotbdotc}
F_{a\dot b \dot c}=\partial_a\,A_{\dot b\dot c}-\partial_{\dot
b}\,A_{a\dot c}+\partial_{\dot c}\,A_{a\dot b}\,,
\ee
\be\label{fdabc}
F_{\dot a\dot b \dot c}=\partial_{\dot a}\,A_{\dot b\dot
c}-\partial_{\dot b}\,A_{\dot a\dot c}+\partial_{\dot c}\,A_{\dot
a\dot b}\,,
\ee
\be\label{tfadot}
{\tilde f}_{a\dot b \dot
c}=\frac{1}{2}\varepsilon_{abc}\,\varepsilon_{\dot b\dot c \dot
a}\,f^{bc\dot a}\,
\ee
and
\be\label{fabdotc}
f_{ab\dot c}=\partial_a\,A_{b\dot c}-\partial_b\,A_{a\dot c}\,.
\ee
Here
$\varepsilon_{abc}$ and $\varepsilon_{\dot a\dot b\dot c}$ are the
antisymmetric unit tensors invariant under $SO(1,2)$ and $SO(3)$,
respectively.

Note that the tensor (\ref{fabdotc}) as well as the Lagrangian
\eqref{hoho} do not contain the components $A_{ab}$ of the gauge
potential. Because of this the Lagrangian \eqref{hoho} is invariant
under the gauge transformations
\be\label{gadb}
\delta\,A_{a\dot b}=\partial_a\,\lambda_{\dot b}-\partial_{\dot
b}\,\lambda_a
\ee
only modulo a total derivative.

As in the case of the formulation of Section \ref{nonc}, eqs.\
(\ref{22}) and (\ref{5}), the $A_{ab}$ component of the chiral field
appears on the mass shell upon integrating out one of the
derivatives of the second order field equations which follow from
the Lagrangian (\ref{hoho}) and have (upon the use of the Bianchi
identities) the form
\cite{Ho1}
\be\label{hohofe}
\frac{\delta S}{\delta A^{a\dot b}}=0\quad \Rightarrow\quad
\partial_{\dot c}\,(F-\tilde f)^{a\dot b \dot c}=0\qquad
\Rightarrow \qquad (F-\tilde f)_{a\dot b \dot c}
=\frac{1}{2}\varepsilon_{\dot b\dot c\dot a}\,\varepsilon_{abc}
\,\partial^{\dot a}\,A^{bc}\,,
\ee
\be\label{hohofe1}
\frac{\delta S}{\delta A^{\dot a\dot b}}=0\quad
\Rightarrow\quad\partial_a\,F^{a\dot b \dot c}+\partial_{\dot a}\,F^{\dot a\dot b
\dot c}=0\,,
\ee
where $A_{ab}(x^\mu)$ is an SO(1,2) antisymmetric tensor field. Then
eq.\  (\ref{hohofe}) takes the form of the duality relation
\be\label{hohofe2}
(F-\tilde F)^{a\dot b \dot c}=0\,\quad \Rightarrow \quad F_{a\dot b
\dot c}=\tilde F_{a\dot b \dot c}\,,
\ee
where
\be\label{tildeF}
\tilde F_{a\dot b \dot
c}\equiv\frac{1}{2}\varepsilon_{abc}\,\varepsilon_{\dot b\dot c \dot
a}\,F^{bc\dot a}\,
\ee
and
\be\label{fabdotc1}
F_{ab\dot c}=f_{ab\dot c}+\partial_{\dot
c}\,A_{ab}=\partial_a\,A_{b\dot c}-\partial_b\,A_{a\dot
c}+\partial_{\dot c}\,A_{ab}
\ee
is a complete gauge invariant $F_{ab\dot c}$ component of the field
strength $F_{\mu\nu\lambda}$.

Substituting $F^{a\dot b\dot c}$
with its dual \eqref{hohofe2}, \eqref{fabdotc1} into
the equation (\ref{hohofe1})  we get
\be\label{fx}
\partial^{\dot a}F_{\dot a\dot b\dot c}+\frac{1}{2}\,\varepsilon_{\dot a\dot b\dot c}
\varepsilon_{abc}\,\partial^{\dot a}\partial^a\,A^{bc} =0,\quad\Rightarrow\quad
F_{\dot a\dot b\dot c}+\frac{1}{2}\,\varepsilon_{\dot a\dot b\dot
c}\,\varepsilon_{abc}\,\partial^a A^{bc}=\varepsilon_{\dot a\dot
b\dot c}\,f(x),
\ee
where $f(x^a)$ is a function of only three coordinates
$x^a=(x^0,x^1,x^2)$, that can always be written as the divergence of
a vector $f(x)=\partial_a\,f^a(x)$. It can thus be absorbed by a
redefinition $A_{ab}\to A_{ab} +\frac{1}{3}\varepsilon_{abc}\,f^c(x)$
without any effect on (\ref{hohofe2}). As a result, eq.\  \eqref{fx}
takes the form of the duality relation
\be
F_{abc}=\frac{1}{6}\,\varepsilon_{abc}
\,\varepsilon_{\dot a\dot b\dot c}\,F^{\dot a\dot b\dot c}\,,
\label{4}
\ee
where
\be\label{abc}
F_{abc}=\partial_a A_{bc}+\partial_b A_{ca}+\partial_c A_{ab}\,
\ee
are components of the field strength of the $D=6$ chiral field which
do not enter the Lagrangian \eqref{hoho}.

Eqs.\ \eqref{hohofe} and (\ref{4}) combine into the $SO(1,5)$
covariant self--duality condition (\ref{selfd}) in which the
components of the $D=6$ antisymmetric tensor
$\varepsilon_{\mu\nu\lambda\rho\sigma\delta}$ are defined as follows
\be\label{63+3}
\varepsilon_{abc\dot a\dot b\dot c}=-\varepsilon_{\dot a\dot b\dot c abc}
=\varepsilon_{a\dot b\dot c bc\dot
a}=\varepsilon_{abc}\,\varepsilon_{\dot a\dot b\dot c}\,.
\ee

\subsection{Symmetries of the non--covariant
formulation}\label{hohohoho} We have already mentioned that the
Lagrangian \eqref{hoho} is invariant under the gauge transformations
\eqref{gadb} only up to a total derivative, because the $A_{ab}$
component of the gauge field does not enter the Lagrangian. We can
restore the complete gauge invariance of the Lagrangian by adding to
it certain terms depending on $A_{ab}$ in such a way that they enter
the Lagrangian as total derivatives and hence do not modify
corresponding equations of motion. With these terms the action takes
the form
\bea
S&=&-\frac14\int d^6x[F_{a\dot b\dot c}(F^{a\dot b\dot c}-\tilde F^{a\dot b\dot c})
+\frac13F_{\dot a\dot b\dot c}(F^{\dot a\dot b\dot c}-
\tilde F^{\dot a\dot b\dot c})]\nn\\
&=&\frac14\int d^6x[\tilde F_{ab\dot c}(\tilde F^{ab\dot c}- F^{ab\dot c})
+\frac13\tilde F_{abc}(\tilde F^{abc}-
 F^{abc})]\,.
\label{S1}
\eea
Since the component $A_{ab}$ enters this action under a total
derivative, in addition to the conventional gauge symmetry
\eqref{gs}, the action (\ref{S1}) is also invariant under the
following local transformations
\be
\delta A_{ab}=\Phi_{ab}(x^\mu)\,,
\label{B}
\ee
which are analogous to the transformations \eqref{6} in Subsection
\ref{nonc}.

We shall now show that, similar to the formulation of Subsection
\ref{nonc}, the action (\ref{S1}) has a non--manifest $D=6$ space--time
symmetry.

By construction, eq.\  (\ref{S1}) is manifestly invariant under the
$SO(1,2)\times SO(3)$ subgroup of the full Lorentz group $SO(1,5)$.
So we should check its invariance under the transformations of the
components of the gauge field $A_{\mu\nu}$ corresponding to the
coset $SO(1,5)/[SO(1,2)\times SO(3)]$ which are parametrized by the
$3\times3$ constant matrix $\lambda^a_{\dot b}$,
\bea\label{cosetL}
\delta_1 A^{a\dot a}&=&\lambda^a_{\dot b} A^{\dot b\dot a}
 +\lambda^b_{\dot c}(x_b\partial^{\dot c}-x^{\dot c}\partial_b)A^{a\dot
 a}\,,\nn\\
\delta_1 A^{\dot a\dot b}&=&-\lambda_a^{\dot a}A^{a\dot b}
 +\lambda_b^{\dot b}A^{b\dot a}
 +\lambda^b_{\dot c}(x_b\partial^{\dot c}-x^{\dot c}\partial_b)A^{\dot a\dot
 b}\,.
\eea
(For simplicity, we work in the gauge $A^{ab}=0$, which can always
be imposed by fixing one of the local symmetries \eqref{gs}, \eqref{B}). The action is
not invariant under the transformations \eqref{cosetL}, but changes as follows
\be
\delta_1 S=-\frac12\int d^6x\, \lambda_b^{\dot c}
(F_{a\dot b\dot c}-\tilde F_{a\dot b\dot c})(F^{ab\dot b}-\tilde F^{ab\dot
b})\,.
\ee
This variation of the action can be compensated if the Lorentz
transformations of the gauge field are accompanied by the following
transformation
\be
\delta_2 A_{a\dot b}=\lambda^c_{\dot d} \,x^{\dot d}(F_{ca\dot b}-\tilde F_{ca\dot
b})\,,\quad
\delta_2 A_{\dot a\dot b}=0\,,\quad(A_{ab}=0)\,.
\label{d2}
\ee
Indeed,
\be
\delta_2 S=\frac12\int d^6x\, \lambda_b^{\dot c}
(F_{a\dot b\dot c}-\tilde F_{a\dot b\dot c})(F^{ab\dot b}-\tilde F^{ab\dot
b})\,.
\ee

As a result, we conclude that the action (\ref{S1}) is invariant
under the following modified $SO(1,5)/[SO(1,2)\times SO(3)]$ transformations
\bea\label{mlorentz}
\delta A^{a\dot a}&=&\lambda^a_{\dot b} A^{\dot b\dot a}
 +\lambda^b_{\dot c}(x_b\partial^{\dot c}-x^{\dot c}\partial_b)A^{a\dot
 a}
+\lambda_c^{\dot d} x_{\dot d}(F^{ca\dot a}-\tilde F^{ca\dot a})
\, ,\nn\\
\delta A^{\dot a\dot b}&=&-\lambda_a^{\dot a}A^{a\dot b}
 +\lambda_b^{\dot b}A^{b\dot a}
 +\lambda^b_{\dot c}(x_b\partial^{\dot c}-x^{\dot c}\partial_b)A^{\dot a\dot
 b}\,,
\label{Lor}
\eea
which together with the $SO(1,2)\times SO(3)$ transformations form a
modified non--manifest $D=6$ Lorentz symmetry of the action
\eqref{S1}. The space--time transformations  become the conventional
$SO(1,6)$ Lorentz transformations on the mass shell, when the gauge
field strength satisfies the self--duality condition.

\subsection{Alternative covariant formulation}\label{covho}
Let us now generalize the action \eqref{S1} in such a way that it
becomes Lorentz-covariant. To this end, by analogy with the
covariant formulation of Section \ref{pst}, we introduce auxiliary
fields which appear in the action in the form of projector matrices
$P_\nu{}^\mu(x)$ and $\Pi_\nu{}^\mu(x)$
\be
P_\nu{}^\rho\,P_\rho{}^\mu=P_\nu{}^\mu(x)\,,\qquad
\Pi_\nu{}^\rho\,\Pi_\rho{}^\mu=\Pi_\nu{}^\mu(x)\,,\qquad
\Pi_\nu{}^\mu=\delta_\nu{}^\mu-P_\nu{}^\mu\,.
\label{P}
\ee
In contrast to the projector \eqref{pa}, we now require that
$P_\nu{}^\mu(x)$ and $\Pi_\nu{}^\mu(x)$ have the rank three and look
for an action that has a local symmetry, analogous to
\eqref{a}, which allows one to gauge fix the projectors to become
the constant matrices
\be\label{gfp}
P_\nu{}^\mu=\left(
\begin{array}{cc}\delta_b{}^a& 0\\
0&0
\end{array}\right)\,,\qquad
\Pi_\nu{}^\mu=\left(
\begin{array}{cc}
0&0\\0&\delta_{\dot b}{}^{\dot a}
\end{array}\right)\,.
\ee
To construct the $SO(1,5)$ covariant generalization of the action
\eqref{S1} we first rewrite it in the form
\be
S=\frac1{4!}\int d^6x[-F_{\mu\nu\rho}F^{\mu\nu\rho} +{\cal
F}_{abc}{\cal F}^{abc} +3{\cal F}_{ab\dot c}{\cal F}^{ab\dot c} ]\,,
\label{S2}
\ee
where
\be
{\cal F}_{\mu\nu\rho}=F_{\mu\nu\rho}-\tilde F_{\mu\nu\rho},\quad
{\cal F}_{abc}=F_{abc}-\tilde F_{abc}\,,\quad {\cal F}_{ab\dot
c}=F_{ab\dot c}-\tilde F_{ab\dot c}\,,\quad \mbox{etc}.
\ee
Note that the field ${\cal F}_{\mu\nu\rho}$ is anti--selfdual,
\be
\tilde {\cal F}_{\mu\nu\rho}=\frac16
\varepsilon_{\mu\nu\rho\alpha\beta\gamma}
{\cal F}^{\alpha\beta\gamma} =-{\cal F}_{\mu\nu\rho}\,.
\ee
Now, using the projectors \eqref{P}, we construct the
Lorentz--covariant generalization of \eqref{S2}
\be
S=\frac1{4!}\int d^6x[-F_{\mu\nu\rho}F^{\mu\nu\rho} +{\cal
F}_{\mu\nu\rho}{\cal F}^{\alpha\beta\gamma} (P^\mu_\alpha
P^\nu_\beta P^\rho_\gamma +3P^\mu_\alpha P^\nu_\beta
\Pi^\rho_\gamma)]\,,
\label{S3}
\ee
or, equivalently,
\bea\label{S31}
S&=&-\frac1{12}\int d^6x\,F_{\mu\nu\rho}{\cal F}^{\alpha\beta\gamma}
\,(\Pi^\mu_\alpha \Pi^\nu_\beta \Pi^\rho_\gamma
+3\Pi^\mu_\alpha \Pi^\nu_\beta P^\rho_\gamma)\nn\\
&=&-\frac1{12}\int d^6x\,\tilde F_{\mu\nu\rho}{\cal F}^{\alpha\beta\gamma}
(P^\mu_\alpha P^\nu_\beta P^\rho_\gamma
+3P^\mu_\alpha P^\nu_\beta \Pi^\rho_\gamma)\,.
\eea

We shall now show that the action \eqref{S3} or \eqref{S31} has
indeed the required local symmetry, provided the projectors are
constructed in an appropriate way from a triplet of scalar fields
$a^r(x)$ $(r=1,2,3)$ being a vector with respect to the $GL(3)$
group. These scalar fields play the same role as the auxiliary field
$a(x)$ of Section \ref{pst}.

\subsection{Symmetries of the covariant action}
Recall that the action (\ref{S1}) is invariant under the local
transformations (\ref{B}). The generalization of this symmetry to
the case of the Lorentz covariant action (\ref{S3}) is
\be
\delta A_{\mu\nu}=P_\mu^\alpha P_\nu^\beta\,
\Phi_{\alpha\beta}(x)\,,\qquad
\delta P_\mu{}^{\nu}=\delta \Pi_\mu{}^{\nu}=0\,.
\label{BB}
\ee
To check this and other symmetries let us perform a general
variation of the action (\ref{S3}) with respect to $A_{\mu\nu}$.
Using the identities
\bea
\varepsilon_{\mu\nu\rho\alpha\beta\gamma}
P^\mu_{\mu'}P^\nu_{\nu'}P^\rho_{\rho'}&=&
-\varepsilon_{\mu\nu\rho\mu'\nu'\rho'}
\Pi^\mu_\alpha \Pi^\nu_\beta \Pi^\rho_\gamma\,,\quad
\varepsilon_{\mu\nu\rho\alpha\beta\gamma}
P^\mu_{[\mu'}P^\nu_{\nu'}\Pi^\rho_{\rho']}=
-\varepsilon_{\mu\nu\rho\mu'\nu'\rho'}
\Pi^\mu_{[\alpha} \Pi^\nu_\beta P^\rho_{\gamma]}\,,\nonumber\\
\\
P_{\mu'}{}^\mu\,P_{\nu'}{}^\nu\,\partial_\lambda\,P_{\mu\nu}&=&0\,,\nonumber
\eea
we find that
\bea
\delta S_{\delta A}&=&\frac1{12}\int d^6x\, \delta F_{\mu\nu\rho}
[-F^{\mu\nu\rho}
+(P^\mu_\alpha P^\nu_\beta P^\rho_\gamma
+3P^\mu_\alpha P^\nu_\beta\Pi^\rho_\gamma
-\Pi^\mu_\alpha\Pi^\nu_\beta\Pi^\rho_\gamma
-3\Pi^\mu_\alpha\Pi^\nu_\beta P^\rho_\gamma){\cal
F}^{\alpha\beta\gamma}]\nn\\
&=&\frac1{12}\int d^6x\, \delta F_{\mu\nu\rho}
[-F^{\mu\nu\rho}
+(-4P^\mu_\alpha P^\nu_\beta P^\rho_\gamma
+6P^\mu_\alpha P^\nu_\beta\delta^\rho_\gamma
-\delta^\mu_\alpha\delta^\nu_\beta\delta^\rho_\gamma)
 {\cal F}^{\alpha\beta\gamma}]\\
&=&\int d^6x\,\delta A_{\mu\nu}[\frac12\partial_\rho{\cal
F}^{\mu\nu\rho} -\frac12 P^\mu_\alpha P^\nu_\beta\partial_\rho{\cal
F}^{\alpha\beta\rho} -(\partial_\rho P^\mu_\alpha)P^\nu_\beta{\cal
F}^{\alpha\beta\rho} -\partial_\rho(P^\rho_\alpha P^\mu_\beta
\Pi^\nu_\gamma\, {\cal F}^{\alpha\beta\gamma})]\,.\nn
\label{45}
\eea
For the variation of $A_{\mu\nu}$ in the form \eqref{BB} we get
\be\label{Phis}
\delta_\Phi S
%=-\int d^6x\, \Phi_{\alpha\beta}{\cal F}^{\nu\sigma\lambda}
 %P^\alpha_\mu P^\beta_\nu \Pi^\rho_\sigma
 %\Pi^\gamma_\lambda(\partial_\rho P^\mu_\gamma)
=-\int d^6x\, \Phi_{\alpha\beta}{\cal F}^{\nu\sigma\lambda}
 P^\alpha_\mu P^\beta_\nu \Pi^\rho_\sigma
 \Pi^\gamma_\lambda(\partial_{[\rho} P^\mu_{\gamma]})\,.
\ee
We see that $\delta_\Phi S=0$ if
\be
\Pi_\sigma{}^\rho\,
 \Pi_\lambda{}^\gamma\,\partial_{[\rho} P_{\gamma]}{}^\mu\,P_{\mu}{}^{\nu}=0\,.
\label{constr}
\ee
Equation (\ref{constr}) is the main differential constraint which
must be satisfied by the projector. It is solved by expressing the
projector in terms of derivatives of a triplet of auxiliary scalar
fields $a^r(x)$ with the index $r=1,2,3$ corresponding to a
3--dimensional representation of $GL(3)$. Namely,
\be\label{3pa}
P_\mu{}^\nu=\partial_\mu a^r \,Y^{-1}_{rs}\,
\partial^\nu a^s\,,\qquad
\Pi_\mu{}^\nu=\delta_\mu{}^\nu-P_\mu{}^\nu\,,
\ee
where $Y^{-1}_{rs}$
%\be\label{M-1}
%Y^{-1}_{rs}\equiv (\partial_\rho a^r \partial^\rho a^s)^{-1}
%\ee
is the inverse matrix for
\footnote{Compare with eq.\  \eqref{pa} of Section \ref{pst}.}
$$
Y^{rs}\equiv\partial_\rho a^r \partial^\rho a^s\,.
$$

Thus, to satisfy the requirement of the local symmetry \eqref{BB},
the projector in the action \eqref{S3} is taken to be in the form
\eqref{3pa}.

In view of the similarity of the structure of the projectors
\eqref{3pa} and \eqref{pa}, one may expect that there is a local
symmetry acting on $a^r(x)$ and $A_{\mu\nu}$, analogous to
\eqref{a}, which allows one to get the gauge condition
\eqref{gfp} by putting
\be\label{agauge}
a^r=\delta_a{}^r\,x^a
\ee
and to recover the modified Lorentz transformation \eqref{d2},
\eqref{mlorentz} of the non--covariant formulation as a compensating
transformation of the local symmetry, preserving the gauge
\eqref{gfp}, \eqref{agauge}.

There is indeed such a local symmetry, \emph{i.e.}
\be
\label{du}
\delta_\varphi \,a^r=\varphi^r(x)\,,\qquad
\delta_\varphi\, A_{\mu\nu}=2\,\varphi^r
\,Y_{rs}^{-1} \,\partial^\gamma a^s{\cal F}_{\alpha\beta\gamma}\,
 P_{[\mu}^\alpha \,\Pi_{\nu]}^\beta\,,
\ee
where $\varphi^r(x)$ are  local parameters. To check the invariance
of the action under \eqref{du} it is also instructive
 to present the variation of the projector
\be\label{dP}
\delta_\varphi\,P_{\mu\nu}=2\,\Pi_{\rho(\mu}\,\partial^\rho \varphi^q
\,Y^{-1}_{qr}\,\partial_{\nu)}a^r\,.
\ee
Note that the variation (\ref{dP}) preserves the constraint
(\ref{constr}), which reflects the fact that the latter is solved by
the projector $P_{\mu}{}^{\nu}$ having the form \eqref{3pa}. A
direct computation shows that the action is invariant under the
variations (\ref{du}) and (\ref{dP}). Indeed,
\bea \label{dS}
  \delta_\varphi\,S = \int d^6x\, T^{\mu\nu}_r \partial_\mu\partial_\nu a^r =
  0\,,
\label{5.18}
\eea
where $T^{\mu\nu}_r$ is the antisymmetric tensor of the form
\be
T^{\mu\nu}_r=-T^{\nu\mu}_r=Y^{-1}_{rs}\,Y^{-1}_{kl}\varphi^k
 \partial^\sigma a^s \partial^\delta a^l
(\Pi^\mu_\alpha\Pi^\nu_\beta\Pi^\rho_\gamma {\cal
F}^{\alpha\beta\gamma}{\cal F}_{\rho\sigma\delta}
+2\,\delta_{[\sigma}^{[\mu}\Pi^{\nu]\rho} {\cal
F}_{\delta]}{}^{\lambda\tau} {\cal
F}_{\rho\alpha\beta}\Pi_\lambda^\alpha P^\beta_\tau)\,.
\ee
The gauge condition (\ref{agauge}) is preserved under the
combined Lorentz transformations and the $\varphi$--transformation
\eqref{du} with parameters $\lambda_{\dot b}^a$ and $\varphi=-\lambda_{\dot
b}^a\,x^{\dot b}$, respectively,
\be
\delta a^r=\delta_L\,a^r+\delta_\varphi a^r=0\,.
\ee
When acting on the components of the gauge field $A_{\mu\nu}$, such
a combined transformation generates the modified Lorentz
transformations (\ref{mlorentz}) of the non--covariant formulation.

\subsection{Coupling to gravity and supersymmetric
generalization}\label{coupling}

Because of the manifest Lorentz covariance of the formulation under
consideration, like in the case of the formulation of
\cite{pst,PST1}, the coupling of the chiral gauge field to
gravity is straightforward. One should only replace in the action
\eqref{S3} and in all the symmetry transformations the Minkowski
metric $\eta_{\mu\nu}$ with a curved $D=6$ metric $g_{\mu\nu}(x)$.
As a result the $D=6$ chiral field action coupled to gravity has the
following form
\be
S=\frac1{24}\int d^6x\sqrt{-g}\,[-F_{\mu\nu\rho}F^{\mu\nu\rho}
+{\cal F}_{\mu\nu\rho}{\cal F}^{\alpha\beta\gamma} (P^\mu_\alpha
P^\nu_\beta P^\rho_\gamma +3P^\mu_\alpha P^\nu_\beta
\Pi^\rho_\gamma)]+\int d^6x\sqrt{-g}\,R\,,
\label{S3g}
\ee
where now the projectors include the $D=6$ metric
\be\label{3pag}
P_\mu{}^\nu=\partial_\mu a^r \,(\partial_\rho a^r
g^{\rho\sigma}\partial_\sigma a^s)^{-1}\,
g^{\nu\lambda}\,\partial_\lambda a^s\,,\qquad
\Pi_\mu{}^\nu=\delta_\mu{}^\nu-P_\mu{}^\nu\,.
\ee

\subsubsection{${\mathcal N}=(1,0)$, $D=6$ tensor supermultiplet}
The simplest ${\mathcal N}=(1,0)$ supersymmetric generalization of
the chiral field action is also straightforward. It involves the
${\mathcal N}=(1,0)$ superpartners of $A_{\mu\nu}$ which are a
scalar field $\phi(x)$ and an $SU(2)$ symplectic Majorana--Weyl
fermion $\psi^{I}_A(x)$ $(A=1,2,3,4;\,I=1,2)$
\cite{Kugo:1982bn,Howe:1983fr}
\be
(\psi^I_A)^*=\bar\psi_{I\dot A}=\varepsilon_{IJ}B_{\dot
A}^B\psi_B^J\,,
\label{MW}
\ee
where the matrix $B$ is unitary and satisfies $B^*B=-1$. The $SU(2)$
indices are raised and lowered according to the following rule
$$
\psi^I=\varepsilon^{IJ}\psi_J\,, \qquad \psi_I=\varepsilon_{IJ}\psi^J\,\qquad
\varepsilon_{12}=-\varepsilon^{12}=1\,.
$$
The existence of the matrix $B$ implies that we do not need spinors
with dotted indices for the fermionic action to be real. To
construct the  ${\mathcal N}=(1,0)$ supersymmetric action one should
just add to the action
\eqref{S3} or
\eqref{S31} the kinetic terms for $\psi_{A I}(x)$ and
$\phi(x)$. The resulting free action is
\bea\label{fermionS}
S&=&\frac1{4!}\int d^6x[-F_{\mu\nu\rho}F^{\mu\nu\rho} +{\cal
F}_{\mu\nu\rho}{\cal F}^{\alpha\beta\gamma} (P^\mu_\alpha
P^\nu_\beta P^\rho_\gamma +3P^\mu_\alpha P^\nu_\beta
\Pi^\rho_\gamma)]\nonumber\\
&-&\frac{1}{2}\int d^6x\,(\psi_I\,\Gamma^\mu\partial_\mu\,\psi^I\,
+\partial_\mu\phi\,\partial^\mu\phi )\,.
\eea
It is invariant under the following supersymmetry transformations
with a constant fermionic parameter $\epsilon^{AI}$
\bea\label{susy}
\delta_\epsilon \,\phi&=&\epsilon^I\,\psi_{I}\,,\nonumber\\
\delta_\epsilon \,
A_{\mu\nu}&=&\epsilon^I\,\Gamma_{\mu\nu}\,\psi_I\,,\nonumber\\
\delta_\epsilon \,\psi_I&=&(\Gamma^\mu\,\partial_\mu\,\phi+\frac{1}{12}\,
\Gamma_{\mu\nu\rho}\,{K}^{\mu\nu\rho})\,\epsilon_I\,,\\
\delta_\epsilon \,a^r(x)&=&0\,,\nonumber
\eea
where
\bea\label{K}
{K}^{\mu\nu\rho}&=& \frac{1}{2} [F^{\mu\nu\rho}+{\tilde
F}^{\mu\nu\rho} +(\Pi^\mu_\alpha\Pi^\nu_\beta\Pi^\rho_\gamma
+6\,\Pi^{[\mu}_\alpha\Pi^\nu_\beta P^{\rho]}_\gamma-P^\mu_\alpha
P^\nu_\beta P^\rho_\gamma -6\,P^{[\mu}_\alpha
P^\nu_\beta\Pi^{\rho]}_\gamma ){\cal
F}^{\alpha\beta\gamma}]\, \nonumber\\
&\equiv& F^{\mu\nu\rho}+(2P^\mu_\alpha P^\nu_\beta P^\rho_\gamma
-6\,P^{[\mu}_\alpha P^\nu_\beta\delta^{\rho]}_\gamma)\,{\cal
F}^{\alpha\beta\gamma}
\eea
is the self--dual tensor ${K}_{\mu\nu\rho}
=\frac{1}{6}\varepsilon_{\mu\nu\rho\alpha\beta\gamma}\,
{K}^{\alpha\beta\gamma}$. The conventions for the $D=6$
gamma--matrices are given in the Appendix.

Note that the supersymmetry transformation (\ref{susy}) of the
fermionic field is unusual. In addition to the field strength
$F^{\mu\nu\rho}$ it contains terms with the anti--self--dual tensor
${\cal F}^{\alpha\beta\gamma}$. On the mass shell,  due to the
self--duality condition ${\cal F}^{\alpha\beta\gamma}=0$, the
supersymmetry variation of the fermions take the conventional form.
Our supersymmetry transformations differ from those given in
\cite{Ho2} (in the linear approximation of their model)
by this additional contribution to the variation of the fermions,
which is required for the supersymmetry of the action.

\subsubsection{${\mathcal N}=(2,0)$,  $D=6$ tensor supermultiplet}

One can combine the supersymmetric action (\ref{fermionS}) with
actions for other matter supermultiplets, e.g. by including into the
model four more scalars and one more Majorana--Weyl spinor and thus
getting the action for an ${\mathcal N}=(2,0)$, $D=6$ chiral tensor
supermultiplet (associated with the physical fields on the M5--brane
worldvolume).

The fields of the ${\mathcal N}=(2,0)$ tensor supermultiplet
transform under the $SO(5)$ R--symmetry of the ${\mathcal N}=(2,0)$
superalgebra as follows. The tensor field is a singlet of $SO(5)$,
the set of the five scalars $\phi^\um$, $\um=1,\ldots,5$ form an
$SO(5)$ vector while the fermions $\psi_{IA}$ carry the index
$I=1,2,3,4$ of a spinor representation of $SO(5)\sim USp(4)$ and the
index $A=1,2,3,4$ of a spinor representation of $SO(1,5)\sim Sp(4)$.
The fermions satisfy the $USp(4)$--symplectic Majorana-Weyl
condition analogous to (\ref{MW})
\be
(\psi^I_A)^*=\bar\psi_{I\dot A}=C_{IJ}B_{\dot A}^B\psi_B^J\,,
\label{MW2}
\ee
where $C_{IJ}$ is a skew--symmetric   $USp(4)$--invariant tensor
\be
C^{IJ}\,C_{JK}=\delta^I_K\,,\qquad C^{IJ}\,C_{IJ}=-4\,,
\ee
which is used to rise and lower the $USp(4)$ indices
\be
\psi^I_A=C^{IJ}\psi_{IA}\,,\qquad
\psi_{IA}=C_{IJ}\psi^J_A\,.
\ee
The anti--symmetric matrices $\gamma^\um_{IJ}=-\gamma^\um_{JI}$
associated with the spinor representation of $SO(5)\sim USp(4)$
satisfy the conventional anti--commutation relations
\be
\gamma^\um_{IJ}\gamma^{\un\,JK}+\gamma^\un_{IJ}\gamma^{\um\,JK}
=2\delta^{\um\un}\delta_I^K\,,
\ee
and the orthogonality and completeness relations
\be
\gamma^\um_{IJ}\gamma^{\un\,IJ}=-4\delta^{\um\un}\,,\quad
\gamma^\um_{IJ}\gamma_\um^{KL}=-2(\delta_I^K\delta_J^L-\delta_I^L\delta_J^K)
-C_{IJ}C^{KL}\,,\quad  C^{IJ}\gamma^\um_{IJ}=0\,.
\ee

The action
\bea
S&=&\frac1{24}\int d^6x[-F_{\mu\nu\rho}F^{\mu\nu\rho}+ {\cal
F}_{\mu\nu\rho}{\cal F}^{\alpha\beta\gamma} (P_\alpha^\mu
P_\beta^\nu P_\gamma^\rho+3P_\alpha^\mu P_\beta^\nu
\Pi_\gamma^\rho)]
\nn\\&&
-\frac12\int d^6x(\psi_{IA}\Gamma^{\mu AB}\partial_\mu \psi^I_B
+\partial_\mu\phi^{\um}\partial^\mu\phi_{\um})
\eea
is invariant under the following ${\mathcal N}=(2,0)$ supersymmetry
variations of the fields
\bea
\delta_\epsilon\phi^{\um}&=&\epsilon^{IA}\gamma^\um_{IJ}\psi^{J}_{A}\,,\\
\delta_\epsilon
A_{\mu\nu}&=&\epsilon^{I}\Gamma_{\mu\nu }\,\psi_{IB}\,,\\
\delta_\epsilon\psi_{IA}&=&(\Gamma^\mu_{AB}\gamma^\um_{IJ}\partial_\mu\phi_{\um}
\epsilon^{JB}
+\frac1{12} (\Gamma_{\mu\nu\rho})_{AB}K^{\mu\nu\rho}\epsilon_I^B)\,,\\
\delta_\epsilon a^r(x)&=&0\,.
\eea
As a further generalization, one can straightforwardly couple the
matter supermultiplets discussed above to supergravity and construct
$D=6$ chiral supergravity actions in a form alternative to that
considered in \cite{Dall'Agata:1997db}--\cite{DePol:2000re}.

\subsection{Comparison of the two actions for the chiral
field}\label{compare}

Let us now compare the chiral field actions of Sections
\ref{chiral} and \ref{hohoh}.
For simplicity, let us consider their
non--covariant versions \eqref{21} and \eqref{S2}. We split the
$SO(5)$ indices $i,j,\ldots$ of the second term of \eqref{21} into
the $SO(3)$ indices $\dot a, \dot b, \ldots$ and $SO(2)$ indices
$I,J=1,2$ and try to rewrite the terms of the action
\eqref{21} in a form in which the indices $I,J$ combine with the
time--like index 0 into the $SO(1,2)$ indices $a,b,c$. As a result,
upon the use of the anti--self--duality of ${\mathcal
F}_{\mu\nu\rho}=(F-\tilde F)_{\mu\nu\rho}$, the action
\eqref{21} can be rewritten in the form
\be\label{212}
S=-\frac{1}{4!}\int d^6x\,[F_{\mu\nu\rho}F^{\mu\nu\rho} -{\cal
F}_{abc}{\cal F}^{abc} -3{\cal F}_{ab\dot c}{\cal F}^{ab\dot c}
+6\,{\mathcal F}_{0\dot a\dot b}\, {\mathcal F}_0{}^{\dot a\dot
b}]\,.
\ee
We see that \eqref{212} differs from the action \eqref{S2} in the
last term which is quadratic in the components ${\mathcal F}_{0\dot
a\dot b}$ of the anti--self--dual part of the field strength. Since
on the mass shell ${\mathcal F}_{\mu\nu\rho}$ vanishes, the two
formulations are classically equivalent, as we have seen in the
previous Sections. It would be of interest to understand whether the
difference of the two chiral--field actions off the mass shell may
lead to different results upon quantization. For instance, the two
formulations may complement each other when the chiral field is
considered in topologically non--trivial backgrounds.

\setcounter{equation}0
\section{Non--linear model for the $D=6$ chiral gauge field from
the BLG action revisited}\label{nonl}

Let us now consider the non--linear chiral field model of
\cite{Ho1,Ho2}. We shall study this model in a simplified case, in which all the
 scalar and spinor matter fields are put to zero, and will show that
the general solution of the field equations of this model results in
the $D=6$ Hodge self--duality of a non--linear field strength of the
chiral field. We shall thus prove the assumption of the authors of
\cite{Ho2} that the field strength is self--dual. The solution of
the equations of motion will allow us to get the dual field strength
components which were missing in \cite{Ho2}, to show that they
transform as scalar fields under the volume preserving
diffeomorphisms and to find a form of the non--linear action of
\cite{Ho2} which only involves components of the field strength and,
hence, is gauge--covariant.

Let us begin with a short overview of the model. It was obtained
from the Bagger--Lambert--Gustavsson model by promoting its
non--Abelian gauge symmetry based on a 3--algebra to an infinite
dimensional local symmetry of volume preserving diffeomorphisms in
an internal 3--dimensional space ${\mathcal N}_3$ whose algebra is
defined by the Nambu bracket
$$ \lbrace f,g,h \rbrace \equiv
\varepsilon^{\dot a\dot b\dot c}\,
\partial_{{\dot a}}\,f\,\partial_{{\dot b}}\,g \,\partial_{{\dot c}}\,h
\,,
$$
where $ f(x^{\dot a})$, $g(x^{\dot a})$ and  $h(x^{\dot a})$ are
functions on ${\mathcal N}_3$, $x^{\dot a}$ are local coordinates of
${\mathcal N}_3$ and $\varepsilon^{\dot a\dot b\dot c}$ is the
$SO(3)$--invariant
 anti--symmetric unit tensor. The 6--dimensional space--time, which
is assumed to be associated with the worldvolume of an M5--brane, is
a fiber bundle with the fiber ${\mathcal N}_3$ over the
3--dimensional space--time of the BLG model. The 6--dimensional
coordinates are $x^\mu=(x^a,x^{\dot a})$ as defined in the previous
Sections.

According to \cite{Ho1,Ho2}, the field content of the
six--dimensional model with the local symmetry of the ${\mathcal
N}_3$--volume preserving diffeomorphisms comprises gauge fields
$A_{a\dot b}(x^\mu)$  and $A^{\dot a}=\frac{1}{2}\,\varepsilon^{\dot
a \dot b \dot c}A_{\dot b\dot c}(x^\mu)$, the five scalar fields
$X^{\underline m}(x^\mu)$, $\underline m=1,\ldots,5$, interpreted as
five bulk directions transversal to the 5--brane worldvolume, and
sixteen fermionic superpartners $\Psi(x^\mu)$ thereof. In what
follows we shall neglect the matter fields $X^{\underline m}$ and
$\Psi$. The fields $A_{a\dot b}$ and $A_{\dot a\dot b}$ are assumed
to be part of the components of the $D=6$ chiral gauge field
$A_{\mu\nu}$ whose components $A_{ab}$ do not appear in the
non--linear model of \cite{Ho1,Ho2}.

The field $A^{\dot a}$ can be combined with the coordinates $x^{\dot
a}$ to form the quantities
\be\label{M1}
X^{\dot a}\equiv \frac{1}{g}\,x^{\dot a}+A^{\dot a}\,(x^\mu)\,,
\ee
where $g$ is a coupling constant. $X^{\dot a}$ are interpreted in
\cite{Ho1,Ho2} as coordinates parametrizing three bulk directions
orthogonal to the M2--branes and parallel to the 5--brane.

A scalar field $\Phi$ and the gauge fields $A_{a\dot b}$ and
$A_{\dot a\dot b}$ transform under local gauge transformations with
parameters $\Lambda_{\dot a}(x^\mu)$ and $\Lambda_{a}(x^\mu)$ as
follows
$$
\delta_{\Lambda}\Phi = g\,\xi^{\dot c}\,\partial_{\dot
c}\Phi\,,
$$
\bea\label{gaugetrans}
\delta_{\Lambda}A_{\dot a \dot b}&=&\partial_{\dot a}\Lambda_{\dot b}
-\partial_{\dot b}\Lambda_{\dot a} +
g\,\xi^{\dot c}\,\partial_{\dot c}A_{\dot a \dot b}\,,\\
\delta_{\Lambda}A_{ a \dot b}&=&\partial_{ a}\Lambda_{\dot b}-\partial_{\dot b}\Lambda_{a} +
g\,\xi^{\dot c}\,\partial_{\dot c}A_{ a \dot b} +g (\partial_{\dot
b}\,\xi^{\dot c})\,A_{a \dot c}\,,
\eea
where
\be\label{dx}
\xi^{\dot a}=-\frac{1}{g}\,\delta_{\Lambda}\, x^{\dot a} = \epsilon^{\dot a\dot b\dot c}\partial_{
\dot b}\,\Lambda_{\dot c}
\ee
so that $ \partial_{\dot a}\,\xi^{\dot a}=\partial_{\dot a}
\epsilon^{\dot a\dot b
\dot c} \partial_{\dot b}\,\Lambda_{\dot c} \equiv 0$, which is the volume preserving
condition.

Here it is worth to mention a subtle point of the construction of
\cite{Ho2}. Namely, the quantities $X^{\dot a}$ defined in
\eqref{M1} transform as scalars under the volume preserving
diffeomorphisms \eqref{gaugetrans}, \eqref{dx}, though they carry
the vector index $\dot a$.  As we shall see below, this property
allows one to construct gauge field strengths which transform as
scalars under
\eqref{gaugetrans} and, hence, can be used to construct a gauge invariant
action of the model within the line of \cite{Ho2}.

If $ \Phi_{i}$  $(i=1,2,3) $ are scalar fields with respect to the
volume preserving diffeomorphisms, their Nambu bracket $\lbrace
\Phi_{1},\Phi_{2},\Phi_{3} \rbrace$ is also a scalar field. This allows one to
define a covariant derivative along the fiber $\mathcal N_3$
\cite{Ho2}
\bea\label{cov1}
 {\mathcal D}_{\dot a}\Phi&=&\frac{g^2}{2}\,\varepsilon_{\dot a \dot b
\dot c} \lbrace
\Phi,X^{\dot b},X^{\dot c}\rbrace
\nonumber\\
&=&[\partial_{\dot a}+g\,(\partial_{\dot b}\,A^{\dot
b}\,\partial_{\dot a}-\partial_{\dot a}\,A^{\dot b}\,\partial_{\dot
b})+\frac{g^2}{2}\,\varepsilon_{\dot a\dot b\dot
c}\,\varepsilon^{\dot d\dot e\dot f}\,\partial_{\dot d}\,A^{\dot b}
\,\partial_{\dot e}\,A^{\dot c}\,\partial_{\dot f}]\Phi\,.
\eea
Note that
$$ {\mathcal D}_{\dot a} X^{\dot b}= \frac{1}{g}\,\delta_{\dot a}{}^{\dot b}\,\det M\, =g^2\,\{X^{\dot
1},\,X^{\dot 2},\,X^{\dot 3}\}\,\delta_{\dot a}{}^{
\dot b}\,,
$$
where
\be\label{M2}
M_{\dot a}{}^{\dot b}=g\,\partial_{\dot a}\,X^{\dot b} =\delta_{\dot
a}^{\dot b}+g\,\partial_{\dot a}\,A^{\dot b}\,.
\ee
One also defines a covariant derivative along the $x^a$ directions
of the $D=6$ space--time which acts on a scalar field $\Phi$ as
follows
\be\label{cov2}
{\mathcal D}_{a}\,\Phi=\partial_a\,\Phi-g\{A_{a\dot b},\,x^{\dot
b},\,\Phi\}=(\partial_a-g\,B_{a}{}^{\dot a}\partial_{\dot a})\,\Phi
\,.
\ee
where
\be\label{b2} B_{a}{}^{\dot a} = \varepsilon^{\dot b \dot c
\dot a}\,\partial_{\dot b}\,A_{a\dot c}\,.
\ee
The definition of the covariant derivative ${\mathcal D}_{a}$ can be
extended to any tensor field $T$ on ${\mathcal N}_3$
\cite{Bandos:2008jv}
\bea\label{cov22}
 {\mathcal D}_{a}\,T&=&(\partial_a-g\,{\mathcal L}_{B_{a}})\,T
\,,
\eea
where ${\mathcal L}_{B_{a}}$ is the Lie derivative along the
${\mathcal N}_3$ vector field $(B_a)^{\dot a}$.

It follows  from \eqref{b2} that $B_{a}{}^{\dot a}$ is a
divergenceless field
\be\label{div}
\partial_{\dot a}\,B_{a}{}^{\dot a} = 0\,,
\ee
which plays the role of the deformation of the Nambu--Poisson
structure when the parameters of the volume preserving
diffeomorphisms depend on $x^a$. Under the gauge transformations
\eqref{gaugetrans}--\eqref{dx}, $B_a{}^{\dot a}$ transforms as
follows
\be\label{dB}
\delta B_a{}^{\dot a}=\partial_a\,\xi^{\dot a}+g\,\xi^{\dot
b}\,\partial_{\dot b}\,B_a{}^{\dot a}-g B_a{}^{\dot
b}\,\partial_{\dot b}\,\xi^{\dot a}\,.
\ee
Therefore, the covariant derivative ${\mathcal D}_a$, eq.
\eqref{cov22}, transforms as a scalar.

Note that, since $X^{\dot a}$ is a scalar under the gauge
transformations \eqref{gaugetrans}--\eqref{dx}, and
$\varepsilon^{\dot a\dot b
\dot c}$ is the invariant tensor, also the covariant derivative
${\mathcal D}_{\dot a}$ is scalar under the gauge transformations
and the matrix $M_{\dot a}{}^{\dot b}$ transforms as a covariant
vector (with respect to the lower index $\dot a$), i.e.
$$
\delta M_{\dot a}{}^{\dot b}=\xi^{\dot c}\,\partial_{\dot
c}\, M_{\dot a}{}^{\dot b}+g\,(\partial_{\dot a}\,\xi^{\dot
c})\,M_{\dot c}{}^{\dot b}\,.
$$
Because of this property the matrix $M_{\dot a}{}^{\dot b}$, as well
as its inverse $M^{-1}_{\dot a}{}^{\dot b}$, acts as a ``bridge"
which converts scalar quantities, like ${\mathcal D}_{\dot a}$, into
vector ones, like $\partial_{\dot a}$, and vice versa. They can also
be regarded as dreibeins which relate global $SO(3)$ vector indices
with ${\mathcal N}_3$ worldvolume indices. For example, the
following useful identity holds for the covariant derivative
\eqref{cov1} acting on a field $\Phi$
\be\label{identity0}
{\mathcal D}_{\dot a}\,\Phi=\det{M}\, M^{-1}_{\dot a}{}^{\dot
b}\,\partial_{\dot b}\,\Phi\,.
\ee
Thus, when $\Phi$ is a scalar field, the above formula demonstrates
how the matrix $M^{-1}_{\dot a}{}^{\dot b}$ transforms the vector
$\partial_{\dot b}\,\Phi$ into the scalar ${\mathcal D}_{\dot
a}\,\Phi$ (with respect to the volume preserving diffeomorphisms).

Note that, as defined in eq. \eqref{cov1}, the derivative ${\mathcal
D}_{\dot a}$ acts covariantly only on the ${\mathcal N}_3$--scalar
fields, but using the matrix $M_{\dot a}{}^{\dot b}$ one can
generalize it to act covariantly also on the ${\mathcal
N}_3$--tensor fields. For instance, the covariant derivative of a
vector field $V_{\dot b}$ is
\be\label{V}
\hat {\mathcal D}_{\dot a}\,V_{\dot b}= {\mathcal D}_{\dot a}\,V_{\dot
b}-({\mathcal D}_{\dot a}\,M_{\dot b}{}^{\dot c})\,M_{\dot
c}{}^{-1\dot d}\,V_{\dot d}\,.
\ee

One can use the covariant derivatives \eqref{cov1} and
\eqref{cov2} to construct covariant field strengths of the gauge fields
$A^{\dot a}$ and $A_{a\dot b}$ as follows
\be\label{H1}
{\mathcal H}_{\dot a\dot b\dot c} + \frac{1}{g}\varepsilon_{\dot a
\dot b \dot c}=\frac{1}{6}\varepsilon_{\dot f [\dot a \dot b} {
\mathcal  D}_{\dot c]}X^{\dot f}
\ee
and
\be\label{H2}
{\mathcal H}_{a \dot a \dot b}=\varepsilon_{\dot a \dot b \dot f}\,{
\mathcal  D}_{ a}\, X^{\dot f}\,.
\ee
Explicitly, the field strengths \eqref{H1} and \eqref{H2} have the
following form
\bea\label{Hho}
{\mathcal H}_{\dot 1\dot 2 \dot 3}&=&\partial_{\dot a}\,A^{\dot
a}+\frac{g}{2}\,(\partial_{\dot a}\,A^{\dot a}\,\partial_{\dot
b}\,A^{\dot b}-\partial_{\dot b}\,A^{\dot a}\,\partial_{\dot
a}\,A^{\dot b})+\frac{g^2}{6}\,\varepsilon_{\dot a \dot b \dot
c}\,\varepsilon^{\dot d \dot f \dot e }\,\partial_{\dot d}\,A^{\dot
a}\,\partial_{\dot f}\,A^{\dot b}\,\partial_{\dot e}\,A^{\dot
c}\,,\\
&\equiv&\frac{1}{g}(\det{M}-1)\,,
\nonumber\\
{\mathcal H}_{a\dot b\dot c}&=&\partial_a\,A_{\dot b\dot
c}-\partial_{\dot b}\,A_{a\dot c}+\partial_{\dot c}\,A_{a\dot
b}-g\,\varepsilon^{\dot d\dot e\dot f}\,\partial_{\dot d}\,A_{a\dot
e}\,\partial_{\dot f}\,A_{\dot b\dot c}\equiv
\varepsilon_{\dot a\dot b\dot c}\,{\mathcal D}_a\,X^{\dot
a}\,.\label{Hho1}
\eea
The field strengths ${\mathcal H}_{\dot a \dot b \dot c}$ and
${\mathcal H}_{\dot a \dot b c}$, which by construction transform as
scalars under the gauge transformations \eqref{gaugetrans}, can also
be derived from the commutator of the covariant derivatives, since
as was shown in
\cite{Ho2}
\be\label{comm1}
[ {\mathcal D}_{\dot a},\,{\mathcal D}_{\dot b}]\,\Phi=-g^2
\lbrace {\mathcal H}_{\dot a \dot b \dot f},\,X^{\dot f},\,\Phi
\rbrace\,,
\ee
\be \label{comm2}
[{\mathcal D}_{ a},\,{\mathcal D}_{\dot b}]\,\Phi=-g^2
\lbrace {\mathcal H}_{a \dot b \dot f},\,X^{\dot f},\,\Phi
\rbrace\,
\ee
and
\be\label{comm3}
[ {\mathcal D}_{ a},\,{\mathcal D}_{ b}]\,\Phi=-\frac{g}{\det
M}\,\varepsilon_{abc}\, {\mathcal D}_d\, \tilde{\mathcal H}^{dc
\dot a}{\mathcal D}_{\dot a}\,\Phi\,.
\ee
 Equation \eqref{comm2}, in which $\Phi$ is taken to be  $X^{\dot
b}$ is nothing but the Bianchi identity
\be\label{bianchi1}
{\mathcal D}_a\tilde {\mathcal H}^{abc}+{\mathcal D}_{\dot
a}\tilde{\mathcal H}^{\dot a  b c}\equiv 0\,,
\ee
where  $\tilde {\mathcal H}^{abc}$ and $\tilde {\mathcal H}^{ab\dot
c}$ are Hodge dual of
\eqref{Hho}, similar to eqs.\  \eqref{tildeF} and \eqref{4} of the
linear case.

In the absence of the scalar and fermion matter fields, the
non--linear chiral field action of \cite{Ho2} has the following form
\be\label{nonla}
S=-\int\, d^6\,x\,\left(\frac{1}{4}\,{\mathcal H}_{a\dot b\dot
c}\,{\mathcal H}^{a\dot b\dot c}+\frac{1}{12}\,{\mathcal H}_{\dot
a\dot b\dot c}\,{\mathcal H}^{\dot a\dot b\dot
c}+\frac{1}{2}\,\varepsilon^{abc}\,B_{a}{}^{\dot
a}\,\partial_{b}\,A_{c\dot a}+ g\,\det B_a{}^{\dot a}\right)\,
\ee
or equivalently (up to a total derivative)
\be\label{nonla1}
S=-\int\, d^6\,x\,\left(\frac{1}{2}\,({\mathcal D}_a
\,X^{\dot b})^2+\frac{g^4}{2}\,\{X^{\dot 1},\,X^{\dot 2},\,X^{\dot 3}\}^2+\frac{1}{2g^2}+\frac{1}{2}\,\varepsilon^{abc}\,B_{a}{}^{\dot
a}\,\partial_{b}\,A_{c\dot a}+ g\,\det B_a{}^{\dot a}\right)\,.
\ee
One can compare the form \eqref{nonla1} of the action (and also the
complete action of \cite{Ho2} including the scalar and the spinor
fields) with the action of the BLG model based on the volume
preserving diffeomorphisms constructed in \cite{Bandos:2008jv}. One
can see that the two actions differ only by the fact that in the
model of
\cite{Ho2} the eight BLG scalars transforming as vectors of an $SO(8)$
R--symmetry are split into 3+5 scalars $X^{\dot a}$ and
$X^{\underline m}$ $(\underline m=1,\cdots,5)$, so that $SO(8)$ is
broken to $SO(3)\times SO(5)$. The scalar fields $X^{\dot a}$ are
identified, via eq. \eqref{M1}, with three directions along
${\mathcal N}_3$ and with components $A_{\dot a\dot b}$ of the
chiral gauge field. Note that both of the models are invariant under
the volume preserving diffeomorphisms, because the above
identification does not change the variation properties of $X^{\dot
a}$, which remain the scalar fields, as discussed above.

The action \eqref{nonla} is invariant under the volume preserving
diffeomorphisms but does not have a covariant form due to the fact
that its last two (Chern--Simons) terms are not expressed in terms
of the field strengths. We shall present a gauge--covariant form of
the action of this model in Subsection
\ref{covac}.

Varying the action \eqref{nonla} with respect to the gauge
potentials $A_{a\dot b}$ and $A_{\dot a\dot b}$ one gets the
covariant equations of motion \cite{Ho2}
\be\label{eofn1}
{\mathcal D}_a\tilde {\mathcal H}^{ab\dot c}+{\mathcal D}_{\dot
a}{\mathcal H}^{\dot a b\dot c}= 0\,,
\ee
\be\label{eofn2}
{\mathcal D}_a {\mathcal H}^{a\dot b\dot c}+{\mathcal D}_{\dot a}
{\mathcal H}^{\dot a \dot b\dot c}= 0\,,
\ee

In \cite{Ho2} the field strength components $\mathcal H_{abc}$ and
$\mathcal H_{ab\dot c}$, which do not show up in the action
\eqref{nonla} and equations of motion \eqref{eofn1} and
\eqref{eofn2}, were not defined, but it was assumed that they are
dual, respectively to \eqref{Hho} and \eqref{Hho1},  so that the
whole non--linear field strength $\mathcal H_{\mu\nu\rho}$ is Hodge
self--dual
\be\label{selfdd}
\mathcal H_{\mu\nu\rho}=\tilde{\mathcal H}_{\mu\nu\rho} \quad \Rightarrow
\quad {\mathcal H}_{a\dot b\dot c}=\frac{1}{2}\varepsilon_{abc}\,\varepsilon_{\dot b\dot c \dot
a}\,{\mathcal H}^{bc\dot a}\,,\qquad  \mathcal H_{\dot a\dot b\dot
c}=-\frac{1}{6}\,\varepsilon_{\dot a\dot b\dot
c}\,\varepsilon^{abc}\,{\mathcal H}_{abc}\,.
\ee
In the next subsection we shall prove this assumption and find the
explicit expressions for $\mathcal H_{ab\dot c}$ and $\mathcal
H_{abc}$ in the following form
\be\label{explicit1}
{\mathcal H}_{ab\dot c}=M^{-1}_{\dot c}{}^{\dot b}\,(F_{ab\dot
b}+{g}\,\varepsilon^{\dot a\dot e\dot k}\,\varepsilon^{\dot d\dot f
\dot g}\,\varepsilon_{\dot k\dot g
\dot b}\,\partial_{\dot a}\,A_{a\dot e}\,
\partial_{\dot d}\,A_{b\dot f})= M^{-1}_{\dot c}{}^{\dot d}\,(F_{ab\dot
d}+{g}\,\varepsilon_{\dot d\dot a \dot b}\,B_{a}{}^{\dot
a}\,B_{b}{}^{\dot b})\,,
\ee
\bea\label{explicit2}
\frac{1}{6}\,\varepsilon^{abc}\,{\mathcal H}_{abc}
&=&\frac{1}{1+\det{M}}\,\left(\frac{1}{3}\,\varepsilon^{abc}\,{F}_{abc}-\frac{g}{2}\,{\mathcal
H}_{a\dot b\dot c}\,{\mathcal H}^{a\dot b\dot
c}-g\,\varepsilon^{abc}\,B_{a}{}^{\dot b}\,F_{bc\dot b}
-4\,g^2\,\det{B_{a}{}^{\dot b}}\right)\nonumber\\
&&\\
&=&\frac{1}{2+\frac{g}{6}\,\varepsilon^{\dot a\dot b\dot
c}\,{\mathcal H}_{\dot a\dot b\dot
c}}\,\left(\frac{1}{3}\,\varepsilon^{abc}\,{F}_{abc}-\frac{g}{2}\,{\mathcal
H}_{a\dot b\dot c}\,{\mathcal H}^{a\dot b\dot
c}-g\,\varepsilon^{abc}\,B_{a}{}^{\dot b}\,F_{bc\dot b}
-4\,g^2\,\det{B_{a}{}^{\dot b}}\right)\,,\nonumber
\eea
where $F_{ab\dot b}$ and $F_{abc}$ are the linear field strengths
\eqref{fabdotc1} and \eqref{abc}, respectively.

By a straightforward calculation one can show that the field
strengths \eqref{explicit1} and
\eqref{explicit2} are covariant and transform as scalars under the
local gauge transformations \eqref{gaugetrans}--\eqref{dx}, as their
dual counterparts \eqref{Hho} and \eqref{Hho1} do. This is achieved
by requiring the following gauge transformations of the potential
$A_{ab}$
\be\label{Aab}
\delta_\Lambda
A_{ab}=\partial_a\,\Lambda_b-\partial_b\,\Lambda_a+g\,\xi^{\dot
b}\,\partial_{\dot b}\,A_{ab}+g\,(A_{a\dot
c}\,\partial_{b}\,\xi^{\dot c}-A_{b\dot c}\,\partial_{a}\,\xi^{\dot
c})\,.
\ee
Note that eq. \eqref{explicit1} is the covariant generalization of a
``pre--field--strength"
\be\label{G}
G_{ab\dot c}=\partial_a\,A_{b\dot c}-\partial_b\,A_{a\dot
c}+g\,\varepsilon_{\dot c\dot a\dot b}\,B_a{}^{\dot a}\,B_b{}^{\dot
b}
\ee
introduced in \cite{Bandos:2008jv}. The addition to \eqref{G} of the
term $\partial_{\dot c}\,A_{ab}$ makes it to transform as a
covariant vector under the gauge transformations
\eqref{gaugetrans}--\eqref{dx} and
\eqref{Aab}, while multiplication by $M^{-1}$ converts this vector
into the gauge scalar ${\mathcal H}_{ab\dot c}$.

\subsection{Solution of the equations of motion and the Bianchi identities}
 Let us now explain how one gets
the field strengths
\eqref{explicit1}, \eqref{explicit2} and the duality relations \eqref{selfdd} by solving
the field equations \eqref{eofn1} and \eqref{eofn2}. The derivation
is similar to that in the linear case of Section 4, but requires more
intermediate steps.

We start with eq.\
\eqref{eofn1} and multiply it by $M^{-1}_{\dot c}{}^{\dot d}$ to get
\be\label{eofn1-1}
M^{-1}_{\dot c}{}^{\dot d}\,{\mathcal D}_a\tilde {\mathcal
H}^{ab\dot c}+M^{-1}_{\dot c}{}^{\dot d}\,{\mathcal D}_{\dot
a}\,{\mathcal H}^{\dot a b\dot c}= 0\,.
\ee
In view of the definition
\eqref{Hho1} of the field strength ${\mathcal H}^{\dot a b\dot c}=-{\mathcal H}^{b \dot a \dot c}$ and the
identity
\eqref{identity0}, the second term of this equation can be written as
a total partial derivative
\bea\label{2nd1}
M^{-1}_{\dot c}{}^{\dot d}{\mathcal D}_{\dot a}{\mathcal H}^{\dot a
b\dot c}&=&
\det{M}\,\varepsilon^{\dot c\dot a \dot f}\,M^{-1}_{\dot c}{}^{\dot d}\,M^{-1}_{\dot a}{}^{\dot
b}\,\partial_{\dot b}\,{\mathcal D}^b\,X_{\dot f}
\nonumber\\
 &=&
\varepsilon^{\dot d\dot b\dot c}\,\partial_{\dot b}(M_{\dot c}{}^{\dot f}\,{\mathcal
D}^b\,X_{\dot f})=\frac{1}{2}\,\varepsilon^{\dot d\dot b\dot
c}\,\partial_{\dot b}(M_{\dot c}{}^{\dot f}\,\varepsilon_{\dot f\dot
a\dot k}\,{\mathcal H}^{b\dot a\dot k})\,.
\eea
The first term of \eqref{eofn1-1} can also be presented as a total
partial derivative
\be\label{1st1}
M^{-1}_{\dot c}{}^{\dot d}\,{\mathcal D}_a\tilde {\mathcal
H}^{ab\dot c}=\varepsilon^{bac}\,M^{-1}_{\dot c}{}^{\dot
d}\,{\mathcal D}_a\,{\mathcal D}_c\,X^{\dot
c}=-\varepsilon^{bac}\,\varepsilon^{\dot a \dot b\dot
d}\,\partial_{\dot a}\,(\partial_a\,A_{c\dot
b}+\frac{g}{2}\,\varepsilon_{\dot b\dot c \dot f}\,B_{a}^{\dot
c}\,B_{c}{}^{\dot f})\,,
\ee
where $ B_{a}^{\dot c}$ is defined in \eqref{b2}.

Substituting \eqref{2nd1} and \eqref{1st1} into eq.\
\eqref{eofn1-1} we get the Bianchi--like equation which, upon taking
off the total derivative (in topologically trivial spaces), produces
the duality relation
\be\label{dual11}
{\mathcal H}^{b \dot a \dot c}= \frac{1}{2}\,\varepsilon^{bcd}\,
\varepsilon^{\dot a\dot
c
\dot b}\,{\mathcal H}_{cd\dot b}\equiv
\tilde {\mathcal H}^{b\dot a \dot c}\,,
\ee
where ${\mathcal H}_{cd\dot b}$ are, by definition, the `$cd\dot
b$'--components of the non--linear gauge field strength given in eq.\
\eqref{explicit1}. The components $A_{ab}$ of the gauge potential
have appeared in $F_{ab\dot b}$ as a result of the integration of
eq.\
\eqref{eofn1-1}. Substituting the above duality relation back into
eq.\
\eqref{eofn1} we get the Bianchi identity
\be\label{bianchi2}
{\mathcal D}_a\tilde {\mathcal H}^{ab\dot c}+{\mathcal D}_{\dot
a}\tilde{\mathcal H}^{\dot a b\dot c}=0\,.
\ee

It is important to observe that the expression \eqref{explicit1} for
${\mathcal H}_{a b \dot c}$ follows directly from the Bianchi
identity
\eqref{bianchi2}, without any need of the equation of motion
\eqref{eofn1}. Indeed, using the identity \eqref{identity0} and the explicit form \eqref{Hho1}
of $\mathcal H_{a\dot b\dot c}$, the Bianchi identity
\eqref{bianchi2} can be rewritten as
\be\label{Bianchibis}
[{\mathcal D}_{a},{\mathcal D}_{ b}]\,X^{\dot c}=-g^2
\lbrace {\mathcal H}_{a  b \dot d},\,X^{\dot d},\,X^{\dot c}
\rbrace\,.
\ee
This expression brings the commutation relation
\eqref{comm3} to the form similar to that of \eqref{comm1} and  \eqref{comm2}.
The explicit form of eq. \eqref{Bianchibis} is
\be
\varepsilon^{\dot a \dot b \dot c}\,\partial_{ \dot a}\left(\partial_{[a}A_{b] \dot b}
+ \varepsilon_{\dot b \dot d \dot f}(B_{a}{}^{\dot a}B_{b}{}^{\dot
b})\right)
\,\partial_{\dot c}\,X^{\dot g} =
\varepsilon^{\dot a \dot b \dot c}\partial_{ \dot a}\left({\mathcal H}_{a b \dot d}
\,\partial_{\dot b}X^{\dot d}\right)\,\partial_{\dot c}\,X^{\dot
g}\,,
\ee
which yields \eqref{explicit1} after integration. Therefore, eq.\
\eqref{explicit1} holds off the mass shell. The equation of motion \eqref{eofn1} together with the Bianchi
\eqref {bianchi2} yields
\be
 {\mathcal D}_{\dot a}( \tilde{\mathcal H}^{\dot a \dot b c} -
{\mathcal H}^{\dot a \dot b c})=0
\ee
that implies the self--duality condition \eqref{dual11}, which was
explicitly shown above.

We can now proceed and solve the second field equation
\eqref{eofn2}. Multiplying it by
$M_{\dot a}{}^{\dot d}\,\varepsilon_{\dot d\dot b\dot c}$ we get
\be\label{eofn2-1}
M_{\dot a}{}^{\dot d}\,\varepsilon_{\dot d\dot b\dot c}\,{\mathcal
D}_a {\mathcal H}^{a\dot b\dot c}+2\,M_{\dot a}{}^{\dot
d}\,{\mathcal D}_{\dot d}\, {\mathcal H}_{\dot 1 \dot 2\dot 3}= 0\,.
\ee
Using the definition \eqref{Hho} of ${\mathcal H}_{\dot a\dot b\dot
c}$ and the identity \eqref{identity0}, one finds that the second
term of this equation is a total derivative
\be\label{2nd2}
2\,M_{\dot a}{}^{\dot d}\,{\mathcal D}_{\dot d}\, {\mathcal H}_{\dot
1 \dot 2\dot 3}=\frac{1}{g}\,\partial_{\dot a}\,((\det M)^2-1)\,,
\ee
where in the r.h.s. we have introduced the unit constant to
ensure that the integral of \eqref{2nd2} does not diverge when $g~
\rightarrow ~0$ and $\det M~\rightarrow~1$.

It now remains to show that also the first term in \eqref{eofn2-1}
is a total derivative modulo the duality relation \eqref{dual11}. To
this end using eqs. \eqref{H2} and
\eqref{explicit1} of ${\mathcal H}_{bc\dot d}$ we rewrite this term in the following form
\bea\label{1st2}
&M_{\dot a}{}^{\dot d}\,\varepsilon_{\dot d\dot b\dot c}\,{\mathcal
D}_a {\mathcal H}^{a\dot b\dot c}=\varepsilon^{abc}\,M_{\dot
a}{}^{\dot d}\,{\mathcal D}_a\,{\mathcal H}_{bc\dot d} +2\,M_{\dot
a}{}^{\dot d}\,{\mathcal D}_a\,
({\mathcal D}^aX_{\dot d}-\frac{1}{2}\,\varepsilon^{abc}{\mathcal H}_{bc\dot d})\,&\nonumber\\
&&\\
&=\varepsilon^{abc}\,{\mathcal D}_a\,(F_{bc\dot
a}+{g}\,\varepsilon_{\dot k\dot g
\dot a}\,B_b{}^{\dot k}\,
B_c{}^{\dot g}) -2g\,({\mathcal D}_a\,\partial_{\dot a}\,X^{\dot
d})\,{\mathcal D}^a\,X_{\dot d}\,+2\,{\mathcal D}_a\,\left(M_{\dot
a}{}^{\dot d}\, ({\mathcal D}^aX_{\dot
d}-\frac{1}{2}\,\varepsilon^{abc}{\mathcal H}_{bc\dot
d})\right)\,.&\nonumber
\eea
Upon some algebra we finally get
\bea\label{1st22}
M_{\dot a}{}^{\dot d}\,\varepsilon_{\dot d\dot b\dot c}\,{\mathcal
D}_a {\mathcal H}^{a\dot b\dot c}&=&\partial_{\dot
a}\,(\varepsilon^{abc}\,\partial_a\,A_{bc}-\frac{g}{2}{\mathcal
H}_{a\dot b\dot c}\,{\mathcal H}^{a\dot b\dot
c}-g\,\varepsilon^{abc}\,B_{a}{}^{\dot b}\,F_{bc\dot b}
-4\,g^2\,\det{B_{a}{}^{\dot b}})\,\nonumber\\
&&+2\,{\mathcal D}_a\,\left(M_{\dot a}{}^{\dot d}\, ({\mathcal
D}^aX_{\dot d}-\frac{1}{2}\,\varepsilon^{abc}{\mathcal H}_{bc\dot
d})\right)\,.
%\equiv \partial_{\dot a}\,{\mathcal H}\,.
\eea
Notice that the first term is a total derivative and the last term
is proportional to the duality relation \eqref{dual11}. Therefore,
when the duality relation \eqref{dual11} is satisfied, eq.\
\eqref{eofn2-1} can be integrated to produce, as in the linear
case of Section \ref{hohoho}, the field strength $\mathcal H_{abc}$
given in
\eqref{explicit2}, the duality relation
\be\label{duality3}
{\mathcal H}_{\dot a\dot b\dot c}=-\frac{1}{6}\,\varepsilon_{\dot
a\dot b\dot c}\,\varepsilon^{abc}\,{\mathcal H}_{abc}
\ee
and the Bianchi identity
\be\label{Bianchi3}
{\mathcal D}_a\,\tilde {\mathcal H}^{a\dot b\dot c}+{\mathcal
D}_{\dot a}\,\tilde {\mathcal H}^{\dot a\dot b\dot c}=0\,.
\ee

One may ask if it is possible to get the expression
\eqref{explicit2} for ${\mathcal H}_{abc}$ starting from the Bianchi
identity
\eqref{Bianchi3} without the use of equations of motion and, in
particular, the duality relation \eqref{dual11}. Unfortunately, for
${\mathcal H}_{abc}$ defined in \eqref{explicit2} this seems not to
be possible. Indeed, if one starts from the Bianchi relation
\eqref{Bianchi3}, adds to it the null term
$$
2\,\det M\,{\partial}_{\dot a}\, {\mathcal H}_{\dot 1
\dot 2\dot 3}-\frac{1}{g}\,\partial_{\dot a}\,((\det M)^2-1)=0\,,
$$
 and repeats the previous calculation
without taking into account the duality condition \eqref{dual11} one
gets
\bea\label{Bianchi4}
&& \partial_{\dot
a}\,\left(\frac{1}{3}\,\varepsilon^{abc}\,F_{abc}-\frac{g}{2}{\mathcal
H}_{a\dot b\dot c}\,{\mathcal H}^{a\dot b\dot
c}-g\,\varepsilon^{abc}\,B_{a}{}^{\dot b}\,F_{bc\dot b}
-4\,g^2\,\det{B_{a}{}^{\dot b}}+\frac{1}{g}\,({\det}^2 M-1)\right)\,\nonumber\\
&&+2\,{\mathcal D}_a\,\left(M_{\dot a}{}^{\dot d}\, ({\mathcal
D}^aX_{\dot d}-\frac{1}{2}\,\varepsilon^{abc}{\mathcal H}_{bc\dot
d})\right)-\frac{1}{3}\det M\,\partial_{\dot
a}\left(\varepsilon^{abc}\,{\mathcal H}_{abc}+\varepsilon^{\dot
a\dot b\dot c}\,{\mathcal H}_{\dot a\dot b\dot c}
\right)=0\,,
\eea
which is satisfied only if one uses the duality relations
\eqref{dual11} and \eqref{duality3}.
Thus we have encountered a peculiar feature of the model under
consideration that if the non--linear ${\mathcal H}_{abc}$ has the
form \eqref{explicit2}, the Bianchi relation \eqref{Bianchi3} is
only satisfied on the mass shell.

\subsection{Gauge--covariant action}\label{covac}

The knowledge of the explicit form
\eqref{explicit2} of ${\mathcal H}_{abc}$  allows us
to rewrite the action
\eqref{nonla} in
the equivalent (modulo total derivatives) but gauge--covariant form
\be\label{nonlacov}
S=-\int\, d^6x\,\left(\frac{1}{8}\,{\mathcal H}_{a\dot b\dot
c}\,{\mathcal H}^{a\dot b\dot c}+\frac{1}{12}\,{\mathcal H}_{\dot
a\dot b\dot c}\,{\mathcal H}^{\dot a\dot b\dot
c}-\frac{1}{144}\,\varepsilon ^{abc}\,{\mathcal H}_{abc}\,{\mathcal
H}_{\dot a\dot b\dot c}\,\varepsilon ^{\dot a\dot b\dot
c}-\frac{1}{12\,g}\,{}\varepsilon ^{abc}\,{\mathcal
H}_{abc}\,\right)\,.
\ee
Note that, as one can check directly, the potential $A_{ab}$ enters
the action \eqref{nonlacov} only under a total derivative and hence
can be dropped out modulo boundary terms. This means that, as in the
case of its free field limit considered in Section \ref{hohoh}, the
action
\eqref{nonlacov} is invariant under the local symmetry \eqref{B}.

Note also that the last term in \eqref{nonlacov} is of a
Chern--Simons type and can be interpreted as a coupling of the
5--brane to the constant background field $C_3$ which has the
non--zero components $C_{\dot a\dot b\dot
c}=\frac{1}{g}\,\varepsilon_{\dot a\dot b\dot c}$ along the $x^{\dot
a}$--directions of the 5--brane. It can thus be rewritten in the
Chern--Simons form similar to that of the M5--brane action (see eq.
\eqref{M5a} below)
$$
\int\,d^6\,x\,\frac{1}{12g}\,\varepsilon^{abc}\,{\mathcal H}_{abc}
=\frac{1}{2}\,\int\,{\mathcal H}_3\wedge\,C_3\,,
$$
where the field strength ${\mathcal H}_3$ and $C_3$ are regarded as
$D=6$ three--forms. The presence of the constant background field
$C_3$ explicitly breaks the $D=6$ Lorentz invariance. It is not
obvious that the action \eqref{nonlacov} can be invariant under a
modified Lorentz symmetry similar to \eqref{mlorentz} of the free
field case. This issue requires additional study.

In the next Section we shall briefly discuss a possibility of the
construction of an alternative non--linear generalization of the
chiral field action \eqref{S3} which may possess (non--manifest)
Lorentz invariance and describe an M5--brane in a generic $D=11$
background.

This completes our consideration of the non--linear chiral field
model. We have proved that the general solution of the non--linear
equations \eqref{eofn1} and
\eqref{eofn2} is amount to the Hodge self--duality of the
non--linear field strength ${\mathcal H}_{\mu\nu\lambda}$. Thus the
number of independent gauge field degrees of freedom of the
non--linear model is the same as in the linear case, {\it i.e.}\
equals to three, as was assumed in
\cite{Ho2}. The knowledge of the explicit form of the field
strengths ${\mathcal H}_{abc}$ and ${\mathcal H}_{ab\dot c}$ has
also allowed us to fined the form \eqref{nonlacov} of the
non--linear action \eqref{nonla} whose Lagrangian is covariant under
the volume preserving diffeomorphisms. We leave for a future the
analysis of the non--linear model in the presence of the scalar and
spinor matter fields.

\setcounter{equation}0
\section{On the relation to the M5--brane}\label{M5}
Let us now briefly discuss the relation of the model of
\cite{Ho1,Ho2} to the known formulations of the M5--brane. In
\cite{Ho2} it was shown that by performing a double dimensional
reduction, the BLG model with the gauge group of $3d$ volume
preserving diffeomorphisms reduces to a five--dimensional
non--commutative $U(1)$ gauge theory with a small non--commutativity
parameter which can be interpreted as an effective worldvolume
theory of a D4--brane in a background with a strong NS--NS gauge
field $B_2$. The symmetries and the fields of these two theories are
known to be related to each other by the Seiberg--Witten map
\cite{Seiberg:1999vs}. Thus, the authors of \cite{Ho2} assumed that the
BLG model with the Nambu--Poisson algebra structure and a week
coupling constant can be related to an M5--brane theory in a $D=11$
background with a constant gauge field $C_3$ (in a strong $C_3$
value limit) and proposed a Seiberg--Witten map relating the two
theories.

M5--branes in a constant $C_3$ field with M2--branes ending on M5
and corresponding non--commutative (quantum) structures have been
considered, {\it e.g.}\ in
\cite{Bergshoeff:2000jn,Michishita:2000hu,Youm:2000kr} using the formulation of
\cite{Howe:1996yn,Howe:1997vn}
and extending the results of \cite{Howe:1997ue} on a self--dual
string soliton on M5. From the perspective of multiple M2--branes
the M5--brane in a constant $C_3$ field was studied in
\cite{Chu:2009iv} making use of a C--field modified Basu--Harvey
equation
\cite{Basu:2004ed}. Recently, in
\cite{Furuuchi:2009zx} these M2--M5 brane systems and corresponding BPS
string solutions on the M5--brane worldvolume have been studied in
the framework of the model of \cite{Ho1,Ho2} in the linear order of
the coupling constant $g$ and an agreement with previous results
have been found via the Sieberg--Witten map.

As we have seen in Section \ref{compare}, at the quadratic order the
alternative actions for the chiral field differ in a term quadratic
in anti--self--dual components of the gauge field strength, so to
study the relation between
\cite{Ho1,Ho2} and the conventional formulations of the M5--branes
in more detail it should be useful to have a Lagrangian formulation
of the M5--brane dynamics in which the components of the field
strength of the chiral gauge field are naturally split into the
$SO(1,2)\times SO(3)$ way, as has been considered in the previous
Sections. Let us briefly discuss how one might construct such a
formulation.

In the known Lagrangian formulation of the M5--brane, the 6d
indices of the chiral field strength are subject to the 1+5
splitting (as has been explained in Section 3). Then the
self--duality condition (\ref{sd0}) or its Lorentz--covariant
counterpart (\ref{sd0l}) gets generalized to a non--linear relation
between the components of the chiral field strength $F_{\mu\nu\rho}$
and its dual $\tilde F_{\mu\nu\rho}$
\cite{M5S}, \cite{PST2}. In the covariant formalism
\cite{PST2} the non--linear self--duality condition has the following Born--Infeld--like
form
\be\label{sdonl}
 {{1}\over{\sqrt{-(\partial a)^2}}}\,H_{\mu\nu\rho}\partial^\rho
a = {{1}\over\sqrt{-\det g}} ~{\delta
\sqrt{\det(g_{\rho\sigma} + i
 \tilde{H}_{\rho\sigma}) }\over{ \delta\tilde{H}_{\mu\nu}} }
\ee
where $g_{\mu\nu}$ is an induced metric on the worldvolume of the
M5--brane, $H_{\mu\nu\rho}\equiv (F+C)_{\mu\nu\rho}$ is the field
strength of the M5--brane worldvolume chiral gauge field
$A_{\mu\nu}$ extended with the worldvolume pullback  of the
antisymmetric gauge field $C_{3}$ of $D=11$ supergravity and
\be\label{tildeH}
\tilde
H_{\mu\nu}={{1}\over{6\sqrt{-(\partial a)^2}}}\,
\partial^\rho
a\,\varepsilon_{\rho\mu\nu\lambda\sigma\tau}
\,H^{\lambda\sigma\tau}\,.
\ee
Eq.\
\eqref{sdonl} follows from the Dirac--Born--Infeld--like M5--brane action
\bea\label{M5a}
S_{{\cal M}_6}&=&
\int_{{\cal M}_6} d^6 x \big[-\sqrt{-\det(g_{\mu\nu} +
i \tilde{H}_{\mu\nu})} -\frac{\sqrt{-g}}{4\,(\partial a)^2}
\partial_\lambda\,a\,{\tilde
H}^{\lambda\mu\nu}H_{\mu\nu\rho}\partial^\rho a\big]
\nonumber\\
 &&
-{1\over 2}\int_{{\cal M}_6}\left[C_6+ H_3 \wedge C_3\right],
\eea
where $C_6$ is the dual of the gauge potential $C_3$. It is
important to notice that the dual field strength $\tilde H_{\mu\nu}$
(\ref{tildeH}) which enters the Born--Infeld part of the action
\eqref{M5a} and the r.h.s. of the self--duality condition
\eqref{sdonl} is invariant under the gauge transformations
\eqref{varm}.

An alternative Lorentz--covariant non--linear self--duality
condition (which does not involve the auxiliary scalar field $a(x)$)
was obtained from the superembedding description of the M5--brane
\cite{Howe:1996yn,Howe:1997fb} which was the first to produce the complete set
of the M5--brane equations of motion
\cite{Howe:1996yn}\footnote{Other cases in which the superembedding condition results in the
dynamical equations of motion include the Type II $D=10$
superstrings and the $D=11$ M2--brane \cite{Bandos:1995zw}, and
D--branes
\cite{Howe:1996mx,Bandos:1997rq}.}. The superembedding self--duality condition is formulated
in terms of a conventional Hodge self--dual rank--3 field
$h_{\mu\nu\rho}=\tilde h_{\mu\nu\rho}$ which is related to the field
strength $H_{\mu\nu\rho}=(F+C)_{\mu\nu\rho}$ by the following
non--linear algebraic equation
\be\label{h}
(F+C)_{\mu\nu\rho}=(m^{-1})_{\mu}{}^{\lambda}\,h_{\lambda\nu\rho}\,,
\qquad h_{\mu\nu\rho}=\tilde h_{\mu\nu\rho}\,,\qquad
F_{\mu\nu\rho}=\partial_{\mu}\,A_{\nu\rho}+\partial_{\nu}\,A_{\rho\mu}+
\partial_{\rho}\,A_{\mu\nu}\,,
\ee
where
\be\label{m}
m_{\mu}{}^{\lambda}=\delta_{\mu}{}^{\lambda}
-2h_{\mu\sigma\nu}\,h^{\sigma\nu\lambda}\,.
\ee
In \cite{Howe:1997vn} it was shown that the non--linear
self--duality condition that follows from the superembedding is
equivalent to the self--duality condition \eqref{sdonl} resulting
from the M5--brane action (more precisely, to its non--covariant
counterpart when the field $a(x)$ is gauge fixed as in \eqref{gfa}).
The relation and the equivalence of the whole systems of the
M5--brane equations of motion which follow from the two alternative
formulations was established in \cite{Bandos:1997gm}.

Yet another derivation of the non--linear self--duality condition
based on its consistency with the M5--brane kappa--symmetry was
given in \cite{Cederwall:1997gg}. This derivation is in a sense
close to the one which follows from the superembedding formulation
since from the point of view of the superembedding the
kappa--symmetry is just a peculiar realization of a conventional
local supersymmetry on the worldvolume of the branes (see
\cite{Sorokin:1999jx} for a review).

The evidence that the two \emph{a priori} different approaches, the
on--shell superembedding formulation \cite{Howe:1996yn} (or its
kappa--symmetric counterpart \cite{Cederwall:1997gg}) and the action
principle of \cite{PST2}, \cite{M5S}, give the equivalent
interrelated descriptions of the classical dynamics of the
M5--brane, points to its uniqueness and, hence, allows one to assume
that any alternative formulation of the M5--brane dynamics should be
related to those described above.

In particular, an appropriate non--linear generalization of the
self--duality conditions \eqref{hohofe2}, \eqref{4} (which would be
alternative to \eqref{sdonl}) should be related to the
Lorentz--covariant superembedding self--duality condition \eqref{h}.
One can try to derive a non--linear self--duality relation
generalizing eqs.\  \eqref{hohofe2} and
\eqref{4} from eq.\  \eqref{h} by performing the (3+3)
splitting of the six--dimensional indices of $H_{\mu\nu\rho}$ and
$h_{\mu\nu\rho}$ and following the lines of ref. \cite{Howe:1997vn}.
The goal is to get these conditions in the following generic form
(whose r.h.s. is invariant under the gauge transformations
\eqref{B} or \eqref{BB})
\be\label{sdnl}
H_{abc}=f_{abc}(\tilde H,\, \tilde H_{d\dot b})\,, \qquad H_{ab\dot
c}=g_{ab\dot c}(\tilde H,\,\tilde H_{d\dot b})\,,
\ee
where $f_{abc}(\tilde H,\, \tilde H_{d\dot b})$ and $g_{ab\dot
c}(\tilde H,\,\tilde H_{d\dot b})$ are tensorial functions of
\be\label{HC}
\tilde H\equiv\frac{1}{6}\varepsilon^{\dot a \dot b\dot c}\,(F+C)_{\dot a \dot
b\dot c}\,,\qquad  \tilde H^d{}_{\dot
b}\equiv\frac{1}{2}\,\varepsilon_{\dot b\dot c\dot d}\,(F+C)^{d\dot
c\dot d}\,.
\ee
Once the explicit form of \eqref{sdnl} is known, one can use it to
construct an M5--brane action in a form alternative to
\eqref{M5a}. Such an action should be invariant under local
symmetries generalizing \eqref{B} and \eqref{mlorentz} (or
\eqref{BB} and \eqref{du}) and should produce the non--linear self--duality
conditions \eqref{sdnl}. Having at hand this alternative formulation
of the M5--brane dynamics one can analyze its relation to the model
of \cite{Ho2} in a limit of a strong constant $C_3$ field. Note that
one cannot directly relate the non--linear self--dual field strength
$h_{\mu\nu\lambda}$  to the self--dual field strength ${\mathcal
H}_{\mu\nu\lambda}$ of the previous Section, since the former is
invariant under the conventional gauge transformations
\eqref{gs}, while the latter is invariant under the gauge
transformations \eqref{gaugetrans}--\eqref{dx}, \eqref{Aab} which
include the volume preserving diffeomorphisms. Therefore, the gauge
field potentials and the field strengths in the two formulations may
only coincide at the free field level when the coupling constat $g$
is set to zero. In the generic case the relation is not
straightforward and can probably be established via a kind of the
Seiberg--Witten map proposed in \cite{Ho2} or by generalizing
results of \cite{Bandos:2008fr}. We leave the study of these
problems for a future research.

\section*{Acknowledgments}
The authors are thankful to Igor Bandos, Kurt Lechner and Linus
Wulff for interest to this work and discussions and Pei--Ming Ho and
Yosuke Imamura for comments on results of refs. \cite{Ho1,Ho2}. I.S.
is grateful to INFN Padova Section and the Department of Physics
``Galileo Galilei'' of Padova University for kind hospitality and
support. Work of P.P., D.S. and M.T. was partially supported by the
INFN Special Initiatives TS11 and TV12. D.S. was partially supported
by an Excellence Grant of Fondazione Cariparo and the grant
FIS2008-1980 of the Spanish Ministry of Science and Innovation. I.S.
also acknowledges partial support from the RFBR grants No
09-02-00078 and No 08-02-90490, the LRSS grant No 2553.2008.2 and
the fellowship of the Dynasty foundation.

\def\theequation{A.\arabic{equation}}
\setcounter{equation}0
\section*{Appendix}
Let $\mu,\nu=0,\ldots,5$ be $SO(1,5)$ Lorentz indices while
$A,B=1,2,3,4$ be the corresponding spinor indices. The matrices
$(\Gamma_\mu)_{AB}$ and $(\Gamma_\mu)^{AB}$ satisfy the Weyl algebra
\be
(\Gamma_\mu)_{AB}(\Gamma_\nu)^{BC}+
(\Gamma_\nu)_{AB}(\Gamma_\mu)^{BC}=-2\eta_{\mu\nu}\delta_A^C
\ee
and are related to each other as follows
\be
(\Gamma_\mu)_{AB}=\frac12\varepsilon_{ABCD}(\Gamma_\mu)^{CD}\,,\qquad
(\Gamma_\mu)^{AB}=\frac12\varepsilon^{ABCD}(\Gamma_\mu)_{CD}\,,
\ee
where $\varepsilon_{1234}=\varepsilon^{1234}=1$.

The $\Gamma$--matrices satisfy the following identities
\be
(\Gamma^\mu)_{AB}(\Gamma_\mu)^{CD}=2(\delta_A^C\delta_B^D-\delta_A^D\delta_B^C)\,,
\quad
(\Gamma_\mu)_{AB}(\Gamma_\nu)^{AB}=4\eta_{\mu\nu}\,,
\quad
(\Gamma_\mu)_{AB}(\Gamma^\mu)_{CD}=2\varepsilon_{ABCD}\,.
\ee

We define the antisymmetrized products of gamma--matrices as
\bea
(\Gamma_{\mu\nu})_A{}^B&=&\frac12[(\Gamma_\mu)_{AC}(\Gamma_\nu)^{CB}
-(\Gamma_\nu)_{AC}(\Gamma_\mu)^{CB}]
=(\Gamma_{[\mu})_{AC}(\Gamma_{\nu]})^{CB}\,,\nn\\
(\Gamma_{\mu\nu\rho})_{AB}&=&
(\Gamma_{[\mu})_{AC}(\Gamma_\nu)^{CD}(\Gamma_{\rho]})_{DB}\,,\nn\\
(\Gamma_{\mu\nu\rho\sigma})_A{}^{B}&=&
(\Gamma_{[\mu})_{AC}(\Gamma_{\nu})^{CD}(\Gamma_\rho)_{DE}
(\Gamma_{\sigma]})^{EB}\,,\quad \mbox{etc.}
\eea
There is the following duality relation for these matrices,
\be
\Gamma_{\mu_1\ldots\mu_k}=-(-1)^{k(k-1)/2}\frac1{(6-k)!}
\varepsilon_{\mu_1\ldots \mu_6}\Gamma^{\mu_{k+1}\ldots\mu_6}\,.
\ee
In particular,
\be
\Gamma_{\alpha\beta}=\frac1{4!}\varepsilon_{\alpha\beta\mu\nu\rho\sigma}
 \Gamma^{\mu\nu\rho\sigma}\,,\quad
\Gamma_{\mu\nu\rho\sigma}=-\frac12\varepsilon_{\mu\nu\rho\sigma\alpha\beta}
\Gamma^{\alpha\beta}\,,\quad
\Gamma_{\mu\nu\rho}=
 \frac16\varepsilon_{\mu\nu\rho\alpha\beta\gamma}\Gamma^{\alpha\beta\gamma}\,.
\ee
One can prove the following identity
\be
\Gamma_{\mu\nu\rho}\Gamma_\sigma
=-(\eta_{\sigma\mu}\Gamma_{\nu\rho}+\eta_{\sigma\nu}\Gamma_{\rho\mu}
+\eta_{\sigma\rho}\Gamma_{\mu\nu})
-\frac12\varepsilon_{\mu\nu\rho\sigma\alpha\beta}
\Gamma^{\alpha\beta}\,.
\ee

%%%%%%%%%%%%%%%%%%%%%%%%%%%%%%%%%%%%%%%%%%%


\begin{thebibliography}{99}
\bibitem{Dirac}
P.~A.~Dirac, ``Quantised singularities in the electromagnetic
field,'' Proc.\ Roy.\ Soc.\ Lond.\ A {\bf 133} (1931) 60;
%%CITATION = PRSLA,A133,60;%%
\\
P.~A.~Dirac, ``The theory of magnetic poles,'' Phys.\ Rev.\  {\bf
74} (1948) 817.
%%CITATION = PHRVA,74,817;%%

\bibitem{Zwan}
D.~Zwanziger, ``Local Lagrangian quantum field theory of electric and
magnetic charges,'' Phys.\ Rev.\ D {\bf 3} (1971) 880.
%%CITATION = PHRVA,D3,880;%%

\bibitem{Deser}
S.~Deser and C.~Teitelboim, ``Duality transformations of abelian and
nonabelian gauge fields,'' Phys.\ Rev.\ D {\bf 13} (1976) 1592.
%%CITATION = PHRVA,D13,1592;%%

\bibitem{FJ}
 R.~Floreanini and R.~Jackiw, ``Selfdual fields as charge density
solitons,'' Phys.\ Rev.\ Lett.\  {\bf 59} (1987) 1873.
%%CITATION = PRLTA,59,1873;%%

\bibitem{HT}
M.~Henneaux and C.~Teitelboim, ``Dynamics of chiral (selfdual) p
forms,'' Phys.\ Lett.\  B {\bf 206} (1988) 650.
%%CITATION = PHLTA,B206,650;%%

\bibitem{tseytlin}
A.~A.~Tseytlin, ``Duality symmetric formulation of string world sheet
dynamics,'' Phys.\ Lett.\ B {\bf 242} (1990) 163;
%%CITATION = PHLTA,B242,163;%%
\\
A.~A.~Tseytlin, ``Duality symmetric closed string theory and
interacting chiral scalars,'' Nucl.\ Phys.\ B {\bf 350} (1991)
395.
%%CITATION = NUPHA,B350,395;%%

\bibitem{SS}
J.~H.~Schwarz and  A.~Sen, ``Duality symmetric actions,'' Nucl.\
Phys.\  B {\bf 411} (1994) 35 [arXiv:hep-th/9304154].
%%CITATION = HEP-TH 9304154;%%

%\cite{Belov:2006jd}
\bibitem{Belov:2006jd}
  D.~Belov and G.~W.~Moore,
  ``Holographic action for the self-dual field,''
  arXiv:hep-th/0605038.
  %%CITATION = HEP-TH/0605038;%%
  \\
%\cite{Belov:2006xj}
%\bibitem{Belov:2006xj}
  D.~M.~Belov and G.~W.~Moore,
  ``Type II actions from 11-dimensional Chern-Simons theories,''
  arXiv:hep-th/0611020.
  %%CITATION = HEP-TH/0611020;%%

\bibitem{mac}
B.~McClain,  F.~Yu and Y.~S.~Wu, ``Covariant quantization of
chiral bosons and Osp(1,1$|$2) symmetry,'' Nucl.\ Phys.\ B {\bf
343} (1990) 689.
%%CITATION = NUPHA,B343,689;%%

\bibitem{wot}
 C.~Wotzasek, ``The Wess-Zumino term for chiral bosons,'' Phys.\
Rev.\ Lett.\ {\bf 66} (1991) 129.
%%CITATION = PRLTA,66,129;%%

\bibitem{ben}
I.~Martin and A.~Restuccia, ``Duality symmetric actions and canonical
quantization,'' Phys.\ Lett.\ B {\bf 323} (1994) 311;
%%CITATION = PHLTA,B323,311;%%
\\
I.~Bengtsson, and A.~Kleppe, ``On chiral p-forms,'' Int.\ J.\ Mod.\
Phys.\  A {\bf 12} (1997) 3397 [arXiv:hep-th/9609102].
%%CITATION = HEP-TH 9609102;%%

\bibitem{berkovits}
N.~Berkovits,  ``Manifest electromagnetic duality in closed
superstring field theory,'' Phys.\ Lett.\ B {\bf 388} (1996)
743 [arXiv:hep-th/9607070].
%%CITATION = HEP-TH 9607070;%%
\\
N.~Berkovits, ``Local actions with electric and magnetic sources,''
Phys.\ Lett.\ B {\bf 395} (1997) 28 [arXiv:hep-th/9610134];
%%CITATION = HEP-TH 9610134;%%
\\
N.~Berkovits, ``Super-Maxwell actions with manifest duality,''
Phys.\ Lett.\ B {\bf 398} (1997) 79 [arXiv:hep-th/9610226].
%%CITATION = HEP-TH 9610226;%%

\bibitem{Siegel}
W.~Siegel, ``Manifest Lorentz invariance sometimes requires
nonlinearity,'' Nucl.\ Phys.\ B {\bf 238} (1984) 307.
%%CITATION = NUPHA,B238,307;%%

\bibitem{KM}
A.~R.~Kavalov and R.~L.~Mkrtchian, ''Lagrangian of the selfduality
equation and D=10, N=2b supergravity,'' Sov.\ J.\ Nucl.\ Phys.\
{\bf 46} (1987) 728.
%%CITATION = SJNCA,46,728.1987\ YAFIA,46,1246;%%

\bibitem{pst}
P.~Pasti, D.~Sorokin and M.~Tonin, ``Note on manifest Lorentz and
general coordinate invariance in duality symmetric models,''
Phys.\ Lett.\ B {\bf 352} (1995) 59 [arXiv:hep-th/9503182];
%%CITATION = HEP-TH 9503182;%%
\\
P.~Pasti,  D.~Sorokin and M.~Tonin, ``Duality symmetric actions with
manifest space-time symmetries,'' Phys.\ Rev.\ D {\bf 52} (1995)
R4277 [arXiv:hep-th/9506109];
%%CITATION = HEP-TH 9506109;%%
\\
P.~Pasti,  D.~Sorokin and M.~Tonin, ``Space-time symmetries in duality
symmetric models,'' [arXiv:hep-th/9509052].
%%CITATION = HEP-TH 9509052;%%

\bibitem{PST1} P.~Pasti, D.~P.~Sorokin, M.~Tonin,
``On Lorentz invariant actions for chiral p-forms,'' Phys.\ Rev.\ D\
{\bf 55} (1997) 6292 [arXiv:hep-th/9611100].
%%CITATION = HEP-TH 9611100;%%
%
%\cite{Maznytsia:1998xw}

\bibitem{Maznytsia:1998xw}
  A.~Maznytsia, C.~R.~Preitschopf and D.~P.~Sorokin,
  ``Duality of self-dual actions,''
  Nucl.\ Phys.\  B {\bf 539} (1999) 438
  [arXiv:hep-th/9805110].
  %%CITATION = NUPHA,B539,438;%%
%\cite{Miao:2000fn}

\bibitem{Miao:2000fn}
  Y.~G.~Miao, R.~Manvelyan and H.~J.~W.~Mueller-Kirsten,
  ``Self-duality beyond chiral p-form actions,''
  Phys.\ Lett.\  B {\bf 482} (2000) 264
  [arXiv:hep-th/0002060].
  %%CITATION = PHLTA,B482,264;%%
\\
%\cite{Miao:2001vx}
%\bibitem{Miao:2001vx}
  Y.~G.~Miao, H.~J.~W.~Muller-Kirsten and D.~K.~Park,
  ``Constructing doubly self-dual chiral p-form actions in D=2(p+1)
  spacetime dimensions,''
  Nucl.\ Phys.\  B {\bf 612} (2001) 215
  [arXiv:hep-th/0106197].
  %%CITATION = NUPHA,B612,215;%%

\bibitem{anomaly}
C.~Imbimbo and A.~Schwimmer, ``The Lagrangian formulation of chiral
scalars,'' Phys.\ Lett.\ B {\bf 193} (1987) 455.
%%CITATION = PHLTA,B193,455;%%
\\
C.~M.~Hull, ``Covariant quantization of chiral bosons and anomaly
cancellation,'' Phys.\ Lett.\ B {\bf 206} (1988) 234.
%%CITATION = PHLTA,B206,234;%%
\\
J.~M.~Labastida and M.~Pernici, ``On the BRST quantization of chiral
bosons,  Nucl.\ Phys.\ B {\bf 297} (1988) 557.
%%CITATION = NUPHA,B297,557;%%
\\
L.~Mezincescu and  R.~I. Nepomechie, ``Critical dimensions for chiral
bosons'', Phys.\ Rev.\ D {\bf 37} (1988) 3067.
%%CITATION = PHRVA,D37,3067;%%

%\cite{Witten:1996hc}
\bibitem{Witten:1996hc}
  E.~Witten,
  ``Five-brane effective action in M-theory,''
  J.\ Geom.\ Phys.\  {\bf 22} (1997) 103
  [arXiv:hep-th/9610234].
  %%CITATION = JGPHE,22,103;%%
\\
 %\cite{Witten:1999vg}
%\bibitem{Witten:1999vg}
E.~Witten,
 ``Duality relations among topological effects in string theory,''
  JHEP {\bf 0005} (2000) 031
  [arXiv:hep-th/9912086].
  %%CITATION = JHEPA,0005,031;%%

\bibitem{nappi}
L.~Dolan and C.~R.~Nappi,  ``A modular invariant partition function
for the fivebrane,''  Nucl.\ Phys.\ B {\bf 530} (1998) 683
[arXiv:hep-th/9806016].
%%CITATION = HEP-TH 9806016;%%

\bibitem{vanden}
C.~Van Den Broeck and K.~Van Hoof, ``Batalin-Vilkovisky gauge-fixing
of a chiral two-form in six dimensions,'' Class.\ Quant.\ Grav.\
{\bf 16} (1999) 4011 [arXiv:hep-th/9905117].
%%CITATION = HEP-TH 9905117;%%


\bibitem{LM}
K.~Lechner, ``Self-dual tensors and gravitational anomalies in 4n+2
dimensions,'' Nucl.\ Phys.\ B {\bf 537} (1999) 361
[arXiv:hep-th/9808025].
%%CITATION = HEP-TH 9808025;%%
\\
K.~Lechner, P.~A.~Marchetti  and M.~Tonin, ``Anomaly free effective
action for the elementary M5-brane,'' Phys.\ Lett.\ B\ {\bf 524}
(2002) 199 [hep-th/0107061].
%%CITATION = HEP-TH 0107061;%%
\\
K.~Lechner  and  P.~Marchetti, ``Dirac branes, characteristic currents
and anomaly cancellations in 5-branes,'' Nucl.\ Phys.\ Proc.\ Suppl.\
{\bf 102} (2001) 94.
%%CITATION = HEP-TH 0103161;%%

\bibitem{cucu}
X.~Bekaert and S.~Cucu,  ``Antifield BRST quantization of
duality-symmetric Maxwell theory,'' JHEP {\bf 0101} (2001) 015
[arXiv:hep-th/0010266].
%%CITATION = HEP-TH 0010266;%%

\bibitem{cima}
O.~M.~Del Cima, O.~Piguet and M.~S.~Sarandy,  ``Finiteness of PST
self-dual models,'' Nucl.\ Phys.\ B {\bf 600} (2001) 387 [arXiv:hep-th/0012067].
%%CITATION = HEP-TH 0012067;%%

\bibitem{bastia}
F.~Bastianelli, S.~Frolov and A.~A. Tseytlin,  ``Conformal anomaly of
(2,0) tensor multiplet in six dimensions and AdS/CFT
correspondence,''
JHEP {\bf 0002} (2000) 013 [arXiv:hep-th/0001041].
%%CITATION = HEP-TH 0001041;%%

%\cite{Berman:2007bv}
\bibitem{Berman:2007bv}
  D.~S.~Berman,
``M-theory branes and their interactions,''
  Phys.\ Rept.\  {\bf 456} (2008) 89,
  arXiv:0710.1707 [hep-th].
  %%CITATION = PRPLC,456,89;%%

\bibitem{Ho1}
P.-M.~Ho, Y.~Matsuo, ``M5 from M2'', JHEP {\bf 0806} (2008) 105,
arXiv:0804.3629 [hep-th].
%%CITATION = ARXIV:0804.3629;%%
%
\bibitem{Ho2} P.-M.~Ho, Y.~Imamura, Y.~Matsuo, S.~Shiba,
``M5-brane in three-form flux and multiple M2-branes,'' JHEP {\bf
0808} (2008) 014, arXiv:0805.2898 [hep-th].
%%CITATION = ARXIV:0805.2898;%%

%\cite{Bagger:2006sk}
\bibitem{Bagger:2006sk}
J.~Bagger and N.~Lambert, ``Modeling multiple M2's,'' Phys.\ Rev.\
D {\bf 75} (2007) 045020 [arXiv:hep-th/0611108];
%%CITATION = PHRVA,D75,045020;%%
\\
%\cite{Bagger:2007jr}
%\bibitem{Bagger:2007jr}
J.A.~Bagger and N.~Lambert, ``Gauge Symmetry and Supersymmetry of
Multiple M2-Branes,'' Phys.\ Rev.\  D {\bf 77} (2008) 065008
arXiv:0711.0955 [hep-th];
%%CITATION = PHRVA,D77,065008;%%
\\
%\cite{Bagger:2007vi}
%\bibitem{Bagger:2007vi}
J.~Bagger and N.~Lambert, ``Comments On Multiple M2-branes,'' JHEP
{\bf 0802} (2008) 105, arXiv:0712.3738 [hep-th].
%%CITATION = JHEPA,0802,105;%%
%\cite{Gustavsson:2007vu}

\bibitem{Gustavsson:2007vu}
A.~Gustavsson, ``Algebraic structures on parallel M2-branes,''
Nucl.\ Phys.\ B\ {\bf 811} (2009) 66,
arXiv:0709.1260 [hep-th].
%%CITATION = ARXIV:0709.1260;%%

%\cite{Bandos:2008jv}
\bibitem{Bandos:2008jv}
  I.~A.~Bandos and P.~K.~Townsend,
 ``SDiff Gauge Theory and the M2 Condensate,''
  JHEP {\bf 0902} (2009) 013,
  arXiv:0808.1583 [hep-th].
  %%CITATION = JHEPA,0902,013;%%


%\cite{Bandos:2008fr}
\bibitem{Bandos:2008fr}
  I.~A.~Bandos and P.~K.~Townsend,
 ``Light-cone M5 and multiple M2-branes,''
  Class.\ Quant.\ Grav.\  {\bf 25} (2008) 245003,
  arXiv:0806.4777 [hep-th].
  %%CITATION = CQGRD,25,245003;%%

%\cite{Bonelli:2008kh}
\bibitem{Bonelli:2008kh}
  G.~Bonelli, A.~Tanzini and M.~Zabzine,
  ``Topological branes, p-algebras and generalized Nahm equations,''
  Phys.\ Lett.\  B {\bf 672} (2009) 390
  [arXiv:0807.5113 [hep-th]].
  %%CITATION = PHLTA,B672,390;%%


%\cite{Chu:2009iv}
\bibitem{Chu:2009iv}
  C.~S.~Chu and D.~J.~Smith,
  ``Towards the Quantum Geometry of the M5-brane in a Constant $C$-Field from
  Multiple Membranes,''
  JHEP {\bf 0904} (2009) 097,
  arXiv:0901.1847 [hep-th].
  %%CITATION = JHEPA,0904,097;%%

\bibitem{PST2} P.~Pasti, D.~P.~Sorokin and M.~Tonin,
``Covariant action for a D=11 five-brane with the chiral field,''
Phys.\ Lett.\ B\ {\bf 398} (1997) 41  [arXiv:hep-th/9701037].
%%CITATION = HEP-TH 9701037;%%
%
\\
%\cite{Bandos:1997ui}
%\bibitem{Bandos:1997ui}
I.~A.~Bandos, K.~Lechner, A.~Nurmagambetov, P.~Pasti, D.~P.~Sorokin
and M.~Tonin, ``Covariant action for the super-five-brane of
M-theory,'' Phys.\ Rev.\ Lett.\  {\bf 78} (1997) 4332 [arXiv:hep-th/9701149];
%%CITATION = HEP-TH 9701149;%%

\bibitem{M5S}
M.~Perry and J.~H. Schwarz, ``Interacting chiral gauge fields in six
dimensions and Born-Infeld  theory,'' Nucl.\ Phys.\ B {\bf 489}
(1997) 47 [arXiv:hep-th/9611065].
%%CITATION = HEP-TH 9611065;%%
\\
J.~H. Schwarz, ``Coupling a self-dual tensor to gravity in six
dimensions,'' Phys.\ Lett.\ B {\bf 395} (1997) 191
[arXiv:hep-th/9701008].
%%CITATION = HEP-TH 9701008;%%
\\
M. Aganagic, J.~Park, C. Popescu and J.~H. Schwarz, ``World-volume
action of the M-theory five-brane,''  Nucl.\ Phys.\ B {\bf 496}
(1997) 191 [arXiv:hep-th/9701166].
%%CITATION = HEP-TH 9701166;%%

%\cite{Bergshoeff:1998vx}
\bibitem{Bergshoeff:1998vx}
  E.~Bergshoeff, D.~P.~Sorokin and P.~K.~Townsend,
``The M5-brane Hamiltonian,''
  Nucl.\ Phys.\  B {\bf 533} (1998) 303
  [arXiv:hep-th/9805065].
  %%CITATION = NUPHA,B533,303;%%


%\cite{Howe:1996yn}
\bibitem{Howe:1996yn}
  P.~S.~Howe and E.~Sezgin,
  ``D=11, p=5,''
  Phys.\ Lett.\  B {\bf 394} (1997) 62
  [arXiv:hep-th/9611008].
  %%CITATION = PHLTA,B394,62;%%

%\cite{Howe:1997fb}
\bibitem{Howe:1997fb}
  P.~S.~Howe, E.~Sezgin and P.~C.~West,
  ``Covariant field equations of the M-theory five-brane,''
  Phys.\ Lett.\  B {\bf 399} (1997) 49
  [arXiv:hep-th/9702008].
  %%CITATION = PHLTA,B399,49;%%

%\cite{Howe:1997vn}
\bibitem{Howe:1997vn}
  P.~S.~Howe, E.~Sezgin and P.~C.~West,
  ``The six-dimensional self-dual tensor,''
  Phys.\ Lett.\  B {\bf 400} (1997) 255
  [arXiv:hep-th/9702111].
  %%CITATION = PHLTA,B400,255;%%

%\cite{Bandos:1997gm}
\bibitem{Bandos:1997gm}
  I.~A.~Bandos, K.~Lechner, A.~Nurmagambetov, P.~Pasti, D.~P.~Sorokin and M.~Tonin,
  ``On the equivalence of different formulations of the M theory  five-brane,''
  Phys.\ Lett.\  B {\bf 408} (1997) 135
  [arXiv:hep-th/9703127].
  %%CITATION = PHLTA,B408,135;%%

%\cite{Cederwall:1997gg}
\bibitem{Cederwall:1997gg}
  M.~Cederwall, B.~E.~W.~Nilsson and P.~Sundell,
``An action for the super-5-brane in D=11 supergravity,''
  JHEP {\bf 9804} (1998) 007
  [arXiv:hep-th/9712059].
  %%CITATION = JHEPA,9804,007;%%

%\cite{Kugo:1982bn}
\bibitem{Kugo:1982bn}
  T.~Kugo and P.~K.~Townsend,
  ``Supersymmetry And The Division Algebras,''
  Nucl.\ Phys.\  B {\bf 221} (1983) 357.
  %%CITATION = NUPHA,B221,357;%%

%\cite{Howe:1983fr}
\bibitem{Howe:1983fr}
  P.~S.~Howe, G.~Sierra and P.~K.~Townsend,
  ``Supersymmetry In Six-Dimensions,''
  Nucl.\ Phys.\  B {\bf 221} (1983) 331.
  %%CITATION = NUPHA,B221,331;%%


%\cite{Dall'Agata:1997db}
\bibitem{Dall'Agata:1997db}
  G.~Dall'Agata, K.~Lechner and M.~Tonin,
  ``Covariant actions for N = 1, D = 6 supergravity theories with  chiral
  bosons,''
  Nucl.\ Phys.\  B {\bf 512} (1998) 179
  [arXiv:hep-th/9710127].
  %%CITATION = NUPHA,B512,179;%%


%\cite{Riccioni:1998pj}
\bibitem{Riccioni:1998pj}
  F.~Riccioni and A.~Sagnotti,
``Self-dual tensors in six-dimensional supergravity,''
  arXiv:hep-th/9812042.
  %%CITATION = HEP-TH/9812042;%%
\\
%\cite{Riccioni:1999xq}
%\bibitem{Riccioni:1999xq}
  F.~Riccioni,
  ``Abelian vector multiplets in six-dimensional supergravity,''
  Phys.\ Lett.\  B {\bf 474} (2000) 79
  [arXiv:hep-th/9910246].
  %%CITATION = PHLTA,B474,79;%%
\\
%\cite{Riccioni:2001bg}
%\bibitem{Riccioni:2001bg}
  F.~Riccioni,
  ``All couplings of minimal six-dimensional supergravity,''
  Nucl.\ Phys.\  B {\bf 605} (2001) 245
  [arXiv:hep-th/0101074].
  %%CITATION = NUPHA,B605,245;%%

%\cite{VanHoof:1999xi}
\bibitem{VanHoof:1999xi}
  K.~Van Hoof,
  ``An action for the (2,0) self-dual tensor multiplet in a conformal
  supergravity background,''
  Class.\ Quant.\ Grav.\  {\bf 17} (2000) 2093
  [arXiv:hep-th/9910175].
  %%CITATION = CQGRD,17,2093;%%

%\cite{DePol:2000re}
\bibitem{DePol:2000re}
  G.~De Pol, H.~Singh and M.~Tonin,
  ``Action with manifest duality for maximally supersymmetric  six-dimensional
  supergravity,''
  Int.\ J.\ Mod.\ Phys.\  A {\bf 15} (2000) 4447
  [arXiv:hep-th/0003106].
  %%CITATION = IMPAE,A15,4447;%%

%\cite{Seiberg:1999vs}
\bibitem{Seiberg:1999vs}
  N.~Seiberg and E.~Witten,
  ``String theory and noncommutative geometry,''
  JHEP {\bf 9909} (1999) 032
  [arXiv:hep-th/9908142].
  %%CITATION = JHEPA,9909,032;%%

%\cite{Bergshoeff:2000jn}
\bibitem{Bergshoeff:2000jn}
  E.~Bergshoeff, D.~S.~Berman, J.~P.~van der Schaar and P.~Sundell,
  ``A noncommutative M-theory five-brane,''
  Nucl.\ Phys.\  B {\bf 590} (2000) 173
  [arXiv:hep-th/0005026].
  %%CITATION = NUPHA,B590,173;%%

%\cite{Michishita:2000hu}
\bibitem{Michishita:2000hu}
  Y.~Michishita,
``The M2-brane soliton on the M5-brane with constant 3-form,''
  JHEP {\bf 0009} (2000) 036
  [arXiv:hep-th/0008247].
  %%CITATION = JHEPA,0009,036;%%

%\cite{Youm:2000kr}
\bibitem{Youm:2000kr}
  D.~Youm,
``BPS solitons in M5-brane worldvolume theory with constant
three-form field,''
  Phys.\ Rev.\  D {\bf 63} (2001) 045004
  [arXiv:hep-th/0009082].
  %%CITATION = PHRVA,D63,045004;%%


%\cite{Howe:1997ue}
\bibitem{Howe:1997ue}
  P.~S.~Howe, N.~D.~Lambert and P.~C.~West,
``The self-dual string soliton,''
  Nucl.\ Phys.\  B {\bf 515} (1998) 203
  [arXiv:hep-th/9709014].
  %%CITATION = NUPHA,B515,203;%%


%\cite{Basu:2004ed}
\bibitem{Basu:2004ed}
  A.~Basu and J.~A.~Harvey,
``The M2-M5 brane system and a generalized Nahm's equation,''
  Nucl.\ Phys.\  B {\bf 713} (2005) 136
  [arXiv:hep-th/0412310].
  %%CITATION = NUPHA,B713,136;%%

%\cite{Furuuchi:2009zx}
\bibitem{Furuuchi:2009zx}
  K.~Furuuchi and T.~Takimi,
``String solitons in the M5-brane worldvolume action with
Nambu-Poisson structure and Seiberg-Witten map'', JHEP {\bf 0908}
(2009) 050, arXiv:0906.3172 [hep-th].
  %%CITATION = ARXIV:0906.3172;%%

%\cite{Bandos:1995zw}
\bibitem{Bandos:1995zw}
  I.~A.~Bandos, D.~P.~Sorokin, M.~Tonin, P.~Pasti and D.~V.~Volkov,
  ``Superstrings and supermembranes in the doubly supersymmetric geometrical
  approach,''
  Nucl.\ Phys.\  B {\bf 446} (1995) 79
  [arXiv:hep-th/9501113].
  %%CITATION = NUPHA,B446,79;%%

%\cite{Howe:1996mx}
\bibitem{Howe:1996mx}
  P.~S.~Howe and E.~Sezgin,
  ``Superbranes,''
  Phys.\ Lett.\  B {\bf 390} (1997) 133
  [arXiv:hep-th/9607227].
  %%CITATION = PHLTA,B390,133;%%

%\cite{Bandos:1997rq}
\bibitem{Bandos:1997rq}
  I.~A.~Bandos, D.~P.~Sorokin and M.~Tonin,
  ``Generalized action principle and superfield equations of motion for  D = 10
  D p-branes,''
  Nucl.\ Phys.\  B {\bf 497} (1997) 275
  [arXiv:hep-th/9701127].
  %%CITATION = NUPHA,B497,275;%%


%\cite{Sorokin:1999jx}
\bibitem{Sorokin:1999jx}
  D.~P.~Sorokin,
  ``Superbranes and superembeddings'',
  Phys.\ Rept.\  {\bf 329} (2000) 1
  [arXiv:hep-th/9906142].
  %%CITATION = PRPLC,329,1;%%

\end{thebibliography}
\end{document}